\documentclass[11pt,draftcls,onecolumn,journal]{IEEEtran}
\usepackage{amsmath}
\usepackage{amssymb}
\usepackage{enumerate,url,multirow}
\usepackage{theorem}
\usepackage{graphicx,color}

\usepackage{pifont}
\usepackage[noadjust]{cite}
\usepackage{citesort}

\usepackage{algorithm,algorithmic}

\def\thetaA{\thetab_{\ab}}

\def\TR{\mathit{TR}}

\def\PM{\kern0pt^{\textrm{{\scriptsize PM}}}\kern0pt}
\def\MMAP{\kern1pt^{\textrm{{\tiny MMAP}}}\kern-1pt}

\def\MAP{^{\textrm{{\tiny MAP}}}}

\def\Ab{{\sbm{A}}\XS}  \def\ab{{\sbm{a}}\XS}
\def\Bb{{\sbm{B}}\XS}  
\def\Cb{{\sbm{C}}\XS}  
\def\Db{{\sbm{D}}\XS}

\def\Gb{{\sbm{G}}\XS}  \def\gb{{\sbm{g}}\XS}
\def\Hb{{\sbm{H}}\XS}  \def\hb{{\sbm{h}}\XS}
\def\Ib{{\sbm{I}}\XS}

\def\Lb{{\sbm{L}}\XS}  
  \def\mb{{\sbm{m}}\XS}

\def\Pb{{\sbm{P}}\XS}  
\def\Qb{{\sbm{Q}}\XS}  \def\qb{{\sbm{q}}\XS}
\def\Rb{{\sbm{R}}\XS}  
\def\Sb{{\sbm{S}}\XS}  
  
\def\Ub{{\sbm{U}}\XS}  
  \def\vb{{\sbm{v}}\XS}
  
\def\Xb{{\sbm{X}}\XS}  
\def\Yb{{\sbm{Y}}\XS}  \def\yb{{\sbm{y}}\XS}
  
\def\mub{{\sbm{\mu}}\XS}
\def\wh#1{\widehat{#1}}                  % Frac en toc
\def\froc#1#2{{#1/#2}}                  % Frac en toc

\def\cc#1{\setlength{\tabcolsep}{0pt}\btabu{c}#1\etabu}
\newcommand{\figc}[2][]
   {\setlength{\tabcolsep}{0pt}\btabu{c}\includegraphics[#1]{#2}\etabu}
\def\yaxis#1{\cc{\rotatebox{90}{{\small #1}}}}

\newcommand{\scalarprod}[2]{{\delta( #1 , #2 )}}
\def\M{^{-1}}
\def\T{^t}
 \def\+{^\dagger}
\def\I{\,|\,}           % "sachant" bien espac\'e pour les formules
\usepackage{afterpage}
\usepackage{textcomp}
\RequirePackage{xspace}
\def\trace{\mathrm{trace}}
\newcommand{\diag}{\ensuremath{\mathrm{diag}}}

\def\x{{\mathbf x}}

\def\hermit{\mbox{\tiny\textsf{H}}}

\newcommand{\RR}{\ensuremath{\mathbb R}}

\theoremstyle{plain}{\theorembodyfont{\rmfamily}%
\theoremstyle{plain}{\theorembodyfont{\rmfamily}%
\renewcommand\theenumi{(\roman{enumi})}
\renewcommand\theenumii{(\alph{enumii})}

\def\qed{\ifmmode\hbox{\hfill\sqb}\else{\ifhmode\unskip\fi%
\nobreak\hfil
\penalty50\hskip1em\null\nobreak\hfil$\blacksquare$
\parfillskip=0pt\finalhyphendemerits=0\endgraf}\fi}

\def\argmax{\mathop{\mathrm{arg\,max}}} % Mieux que \def\argmax{\arg\max}
\RequirePackage{amsmath}
\RequirePackage{xspace}
\RequirePackage{bbm}

\def\XS{\xspace}
\DeclareMathAlphabet{\mathb}{OML}{cmm}{b}{it}
\def\sbm#1{\ensuremath{\mathb{#1}}}                % Style gras italique (necessite amsmath)           
\def\sbmm#1{\ensuremath{\boldsymbol{#1}}}          % Style gras italique (necessite amsmath)           
\def\sdm#1{\ensuremath{\mathrm{#1}}}               % Style droit en math
\def\sbv#1{\ensuremath{\mathbf{#1}}}               % Style gras droit
\def\scu#1{\ensuremath{\mathcal{#1\XS}}}           % Style cursif
\def\scb#1{\ensuremath{\boldsymbol{\mathcal{#1}}}} % Style gras cursif
\def\sbl#1{\ensuremath{\mathbbm{#1}}}              % Style blackboard (necessite bbm)
\def\zerob   {{\sbv{0}}\XS}

\def\Nc{{\scu{N}}\XS}   
\def\pth#1{\left(#1\right)}                
\def\acc#1{\left\{#1\right\}}              
\def\cro#1{\left[#1\right]}                
               
\def\norm#1{\left\|#1\right\|}             

\def\bigpth#1{\bigl(#1\bigr)}              \def\biggpth#1{\biggl(#1\biggr)}
\def\bigacc#1{\bigl\{#1\bigr\}}            
\def\bigcro#1{\bigl[#1\bigr]}

\def\Bigpth#1{\Bigl(#1\Bigr)}

\newcommand{\DQe}{\mathcal{D}} % Ensemble des probabilites sur Z
\def\Ab{{\sbm{A}}\XS}  \def\ab{{\sbm{a}}\XS}
\def\Bb{{\sbm{B}}\XS}  
\def\Cb{{\sbm{C}}\XS}  
\def\Db{{\sbm{D}}\XS}

\def\Gb{{\sbm{G}}\XS}  \def\gb{{\sbm{g}}\XS}
\def\Hb{{\sbm{H}}\XS}  \def\hb{{\sbm{h}}\XS}
\def\Ib{{\sbm{I}}\XS}

\def\Lb{{\sbm{L}}\XS}  
  
\def\mb{{\sbm{m}}\XS}

\def\Pb{{\sbm{P}}\XS}  
\def\Qb{{\sbm{Q}}\XS}  \def\qb{{\sbm{q}}\XS}
\def\Rb{{\sbm{R}}\XS}  
\def\Sb{{\sbm{S}}\XS}  
  
\def\Ub{{\sbm{U}}\XS}  
  \def\vb{{\sbm{v}}\XS}
  
\def\Xb{{\sbm{X}}\XS}  
\def\Yb{{\sbm{Y}}\XS}  \def\yb{{\sbm{y}}\XS}
  
\def\ellb{{\sbm{\ell}}\XS}  %\def\\varepsilonb{{\sbm{\varepsilon}}\XS}
\def\th{-th\XS}
      \def\Sigmab    {{\sbmm{\Sigma}}\XS}
\def\Espud#1#2{{\mathrm{E}}_{#1}\bigcro{#2}}
\def\wt#1{\widetilde{#1}} 
\def\Hb{{\sbm{H}}\XS}  \def\hb{{\sbm{h}}\XS}
\def\M{^{-1}} 
  
\def\tD{{\sdm{t}}\XS}
\def\T{^\tD} \def\+{^\dagger}

\newcommand{\Qe}{\mathcal{Q}} % Espace d'etat du processus cach\'e.
\newcommand{\He}{\mathcal{H}} % Espace d'etat du processus cach\'e.
\newcommand{\Ae}{\mathcal{A}} % Espace d'etat du processus cach\'e.
\def\Dc{{\scu{D}}\XS}   
   
\def\Fc{{\scu{F}}\XS}   
\def\Gc{{\scu{G}}\XS}

\def\Nc{{\scu{N}}\XS}   
   
\def\Pc{{\scu{P}}\XS}   \def\Pcb{{\scb{P}}\XS}

\def\MAP{^{\kern1pt{\rm MAP}\kern-1pt}}
\def\WITH{\quad \text{with} \quad}

\def\widebar#1{\overline{#1}}
\usepackage{color,soul}

\def\Cbb{{\sbl{C}}\XS}  
  
\def\Ebb{{\sbl{E}}\XS}

\def\Nbb{{\sbl{N}}\XS}

\def\Qbb{{\sbl{Q}}\XS}  
\def\Rbb{{\sbl{R}}\XS}

\def\Zbb{{\sbl{Z}}\XS}

\def\alphab      {{\sbmm{\alpha}}\XS}
\def\betab       {{\sbmm{\beta}}\XS}
      
\def\Gammab    {{\sbmm{\Gamma}}\XS}
      \def\Deltab    {{\sbmm{\Delta}}\XS}

\def\varepsilonb {{\sbmm{\varepsilon}}\XS}

\def\thetab      {{\sbmm{\theta}}\XS}      \def\Thetab    {{\sbmm{\Theta}}\XS}

    \def\Lambdab   {{\sbmm{\Lambda}}\XS}
\def\mub         {{\sbmm{\mu}}\XS}

      \def\Sigmab    {{\sbmm{\Sigma}}\XS}

\def\PM{\kern0pt^{\textrm{{\scriptsize PM}}}\kern0pt}
\def\MMAP{\kern1pt^{\textrm{{\tiny MMAP}}}\kern-1pt} 

\def\dd{\,d}                            % doit etre en italique en anglais
\def\I{\,|\,}           % "sachant" bien espac\'e pour les formules
\def\rem#1{}                    % rem{bla bla bla}
\def\mSENSE{\texttt{mSENSE}\XS}

\usepackage{xcolor}
\usepackage[normalem]{ulem}

\definecolor{indigo}{rgb}{.29,0.,0.51}

\usepackage{cite}

\input{alphabet}
\input{abrege}
\def\pth#1{\left(#1\right)}                
\def\acc#1{\left\{#1\right\}}              
\def\cro#1{\left[#1\right]}                
               
\def\norm#1{\left\|#1\right\|}

\def\bigpth#1{\bigl(#1\bigr)}              \def\biggpth#1{\biggl(#1\biggr)}
\def\bigacc#1{\bigl\{#1\bigr\}}            
\def\bigcro#1{\bigl[#1\bigr]}

\def\Bigpth#1{\Bigl(#1\Bigr)}

\def\diag{{\mathrm{diag}}}

							\def\Espud#1#2{{\mathrm{E}}_{#1}\bigcro{#2}}

\def\th{{\mathrm{th}}}                % th hyperbolique en FR
\def\WITH{\text{with\:}}

\def\arrayp{\renewcommand{\arraystretch}{.7}\setlength{\arraycolsep}{2pt}}
\def\tabp{\renewcommand{\arraystretch}{.7}\setlength{\tabcolsep}{2pt}}

\newsavebox{\fminibox}
\newlength{\fminilength}
\newenvironment{fminipage}[1][\linewidth]
  {\setlength{\fminilength}{#1}%-2\fboxsep-2\fboxrule}%
   \begin{lrbox}{\fminibox}\begin{minipage}{\fminilength}}
  {\end{minipage}\end{lrbox}\noindent\fbox{\usebox{\fminibox}}}

\def\M{^{-1}} \def\T{^\tD} \def\+{^\dagger}
\def\I{\,|\,}           % "sachant" bien espac\'e pour les formules
\def\nequiv{\not\kern-.05em\equiv}
\def\egal{\kern-.5em=\kern-.5em}        % Moins d'espace autour de "="
\def\propt{\kern-.2em\propto\kern-.2em} % Idem
\def\wh#1{\widehat{#1}}                 % Sombrero !
\def\wt#1{\widetilde{#1}} 

\def\argmax{\mathop{\mathrm{arg\,max}}} % Mieux que \def\argmax{\arg\max}
\def\froc#1#2{{#1/#2}}                  % Frac en toc

\def\dd{\,d}                            % doit etre en italique en anglais

\def\intdouble{\int\kern-0.3em\int}
\def\inttriple{\int\kern-0.3em\int\kern-0.3em\int}

\def\rond#1{\overset{\kern-0.33em~_\circ}{#1}}
\def\rondit[#1]#2{\overset{\kern#1~_\circ}{#2}}

\def\babs{\begin{abstract}}             \def\eabs{\end{abstract}}
\def\barr{\begin{array}}                \def\earr{\end{array}}
\def\bcc{\begin{center}}                \def\ecc{\end{center}}
\def\cl#1{\centerline{#1}}
\def\bdes{\begin{description}}          \def\edes{\end{description}}
\def\bdoc{
\begin{document}}             \def\edoc{\end{document}}
\def\ben{\begin{enumerate}}             \def\een{\end{enumerate}}
\def\beqn{\begin{eqnarray}}             \def\eeqn{\end{eqnarray}}
\def\beqnl#1{\beqn\label{#1}}           \def\eeqnl#1{\label{#1}\eeqn}
\def\beqnx{\begin{eqnarray*}}           \def\eeqnx{\end{eqnarray*}}
\def\bseqn{\begin{subeqnarray}}         \def\eseqn{\end{subeqnarray}}
\def\beq#1\eeq{\begin{equation}#1\end{equation}}
\def\bal#1\eal{\begin{align}#1\end{align}}
\def\balx#1\ealx{\begin{align*}#1\end{align*}}
\def\beqx{$$}                           \def\eeqx{$$}
\def\bfig{\protect\begin{figure}}       \def\efig{\protect\end{figure}}
\def\bfigx{\protect\begin{figure*}}     \def\efigx{\protect\end{figure*}}
\def\bfigt{\protect\begin{figurette}}   \def\efigt{\protect\end{figurette}}
\def\bfl{\begin{flushleft}}             \def\efl{\end{flushleft}}
\def\bfr{\begin{flushright}}            \def\efr{\end{flushright}}
\def\bit{\begin{itemize}}               \def\eit{\end{itemize}}
\def\bmi{\begin{minipage}}              \def\emi{\end{minipage}}
\def\bfmi{\begin{fminipage}}            \def\efmi{\end{fminipage}}
\def\bpic{\begin{picture}}              \def\epic{\end{picture}}
\def\bqu{\begin{quote}}                 \def\equ{\end{quote}}
\def\bqun{\begin{quotation}}            \def\equn{\end{quotation}}
\def\bsl{\begin{slide}}                 \def\esl{\end{slide}}
\def\btabb{\begin{tabbing}}             \def\etabb{\end{tabbing}}
\def\btabl{\begin{table}}               \def\etabl{\end{table}}
\def\btablx{\begin{table*}}             \def\etablx{\end{table*}}
\def\btab{\begin{tabular}} %\def\etab{\end{tabular}} Conflit avec eta gras... Elle est pas grasse, ma soeur !
\def\btabu{\begin{tabular}}             \def\etabu{\end{tabular}}
\def\btabx{\begin{tabular*}}            \def\etabx{\end{tabular*}}
\def\bbib{\begin{thebibliography}{999}} \def\ebib{\end{thebibliography}}
\def\bver{\begin{verbatim}}             \def\ever{\end{verbatim}}
\def\bca{\begin{cases}}                          \def\eca{\end{cases}}

% TEX 7(ascii) bits
%
% TRANSDEF.tex           LaTeX document
% Author: JI/PC          Date  : 07-08-99
% 
% D\'EFINITIONS POUR TRANSPARENTS
%

\typeout{\space}
\typeout{\space\space\space\space Fichier 'transdef.tex' -- SHFJ 2001}
\typeout{\space\space\space\space (necessite le color)}
\typeout{\space}

\def\hspd{\hspace*{1cm}}

\def\fleche{{\Large \raisebox{-.05cm}{$\Rightarrow$}}}
\def\carre{\setlength{\shadowsize}{2pt}\raisebox{-.05cm}{\shadowbox{}}}
\def\flechedbl{{\Large \raisebox{-.05cm}{$\Rightarrow$}}}
\def\tiret{{---~~ }}

\def\B{\mbox{\large$\bullet$}}
\def\C{\mbox{\large$\circ$}}
\def\X{\mbox{\large$\times$}}
\def\x{\raisebox{-3pt}{\Huge$\times$}}
\def\cc#1{\setlength{\tabcolsep}{0pt}\btabu{c}#1\etabu}%ji
\def\fleche{{\Large \raisebox{-.05cm}{$\Rightarrow$}}\XS}
\def\fleched{\XS{\Large \raisebox{-.05cm}{$\Leftrightarrow$}}\XS}
\def\fleches{{\Large \raisebox{-.05cm}{$\rightarrow$}}\XS}
\def\dfleches{{\Large \raisebox{-.05cm}{$\leftrightarrow$}}\XS}
\def\upfleche{{\Large \raisebox{-.05cm}{$\uparrow$}}\XS}
\def\Upfleche{{\Large \raisebox{-.05cm}{$\Uparrow$}}\XS}
\def\dofleche{{\Large \raisebox{-.05cm}{$\downarrow$}}\XS}
\def\Dofleche{{\Large \raisebox{-.05cm}{$\Downarrow$}}\XS}
\def\sefleche{{\Large \raisebox{-.05cm}{$\searrow$}}\XS}
\def\swfleche{{\Large \raisebox{-.05cm}{$\swarrow$}}\XS}
\def\tiret{{---~~ }}
\def\Boulet#1{\hsp{\B~#1}}
\def\Bouletdec#1{\hspd{\B~#1}}
\def\Cdec#1{\hspd{\C~#1}}

\def\cl#1{\centerline{#1}}
\def\eqcl#1{\cl{$\displaystyle#1$}}
\def\eq#1{{$\displaystyle#1$}}
\def\cites#1{{\small\cite{#1}}}
\def\arrayp{\renewcommand{\arraystretch}{.7}\setlength{\arraycolsep}{2pt}}
\def\tabp{\renewcommand{\arraystretch}{.7}\setlength{\tabcolsep}{2pt}}

         % conventions d'espace pour transparents
\def\sk{\vskip.1in}
\def\hsp{\hspace*{.5cm}}
\def\hspp{\hspace*{1cm}}
\def\vsps{\vspace*{.6cm}}
\def\vspc{\vspace*{.5cm}}
\def\vsp{\vspace*{.5cm}}
\def\hspn{\hspace*{-.5cm}}
\def\hspnu{\hspace*{-.4cm}}
\def\vspq{\vspace*{.4cm}}
\def\vspd{\vspace*{.2cm}}
\def\vspt{\vspace*{.3cm}}
\def\vspu{\vspace*{.1cm}}
\def\vspn{\vspace*{-.1cm}}
\def\vspnd{\vspace*{-.2cm}}
\def\vspnq{\vspace*{-.45cm}}

%%%%%%%%%%%%%%%%%%%%%%%%%%%%%%%%%%%%%%%
        
% Eq belles pour transparents
\def\eqcl#1{\cl{$\displaystyle#1$}}
\def\eq#1{{$\displaystyle#1$}}
\def\sump{\mathop{\raisebox{-.3ex}{\text{\Large$\Sigma$}}}}
\def\sumpp{\mathop{\raisebox{-.2ex}{\text{\large$\Sigma$}}}}
\def\prodp{\mathop{\raisebox{-.3ex}{\text{\Large$\Pi$}}}}
\def\prodpp{\mathop{\raisebox{-.2ex}{\text{\large$\Pi$}}}}

% Les points importants sur 1 transparent
\def\P{\mbox{\raisebox{5pt}{\scriptsize($P$)}}}
\def\diagou{\mbox{\Large$\diagup$}}
\def\diagod{\mbox{\Large$\diagdown$}}

% Citations belles pour transparents
\def\cites#1{{\small\cite{#1}}}
\def\citesa#1{{\small\cite[\kern-.3em a]{#1}}}

\RequirePackage{color}

% En gris, sserif
%\def\titresl#1{\setlength{\shadowsize}{4pt}\cl{\shadowbox{\mbox{~~\sc #1~~}}}}
\newcommand{\titresl}[2][]{\cl{
\psframebox*[shadow=true,shadowsize=4pt,linestyle=none,fillcolor=lightgray]{~~\uppercase{#1{\footnotesize #2}}~~}}}

% Avec couleur, sserif
%\newcommand{\titresl}[2][]{\setlength{\shadowsize}{4pt}\setlength{\fboxsep}{0pt}
%           \cl{\shadowbox{\psframebox*[fillcolor=lightyellow]{~~\uppercase{#1{\footnotesize #2}}~~}\kern-1pt}}}
% Sans couleur, sserif
%\newcommand{\titresl}[2][]{\setlength{\shadowsize}{4pt}\setlength{\fboxsep}{.5pt}
%                           \cl{\shadowbox{\psframebox*{~~\uppercase{#1{\footnotesize #2}}~~}}}}
%                           \cl{\shadowbox{\psframebox*{~~\fcs{#1{ #2}}~~}}}}
% Avec couleur, avec serif
%\newcommand{\titresl}[2][]{\setlength{\shadowsize}{4pt}\setlength{\fboxsep}{.5pt}
%                           \cl{\shadowbox{\psframebox*[fillcolor=lightyellow]{~~\textsc{#1#2}~~}}}}
% Sans couleur, avec serif
%\newcommand{\titresl}[2][]{\setlength{\shadowsize}{4pt}\setlength{\fboxsep}{.5pt}
%                           \cl{\shadowbox{\psframebox*{~~\textsc{\red#1#2}~~}}}}
\definecolor{midgray}{rgb}{.5,.5,.5}

\def\carregris{\setlength{\shadowsize}{2pt}\setlength{\fboxsep}{0pt}\setlength{\fboxrule}{0pt}\raisebox{-.05cm}{\shadowbox{{\kern-.5pt \color{midgray}$\blacksquare$\kern-1pt}}}}
\def\Carre#1{\hsp{\carregris}~~\textbf{#1}}
\def\Annee#1#2#3{{\color{#2} #1}#3}

\def\carrerouge{\setlength{\shadowsize}{2pt}\setlength{\fboxsep}{0pt}\setlength{\fboxrule}{0pt}\raisebox{-.05cm}{\shadowbox{{\kern-.5pt \color{RougePhil}$\blacksquare$\kern-1pt}}}}

\def\carrebleu{\setlength{\shadowsize}{2pt}\setlength{\fboxsep}{0pt}\setlength{\fboxrule}{0pt}\raisebox{-.05cm}{\shadowbox{{\kern-.5pt \color{BleuPhil}$\blacksquare$\kern-1pt}}}}

\def\Carreb#1{\hsp{\carrebleu}~~\textbf{#1}}
\def\Carrer#1{\hsp{\carrerouge}~~\textbf{#1}}

\def\Carred#1#2{\hspace*{#2cm}{\carregris~~\textbf{#1}}}

\def\incircgris#1{\raisebox{.25mm}{\pscirclebox[framesep=1pt,fillstyle=solid,fillcolor=lightgray,linestyle=none]{\textbf{\white\footnotesize#1}}}}
              \def\incircgrisfonc#1{\raisebox{.25mm}{\pscirclebox[framesep=1pt,fillstyle=solid,fillcolor=darkgray,linestyle=none]{\textbf{\white\scriptsize#1}}}}

 \def\Cgrdec#1{\hspd{{\color{midgray}\B}~#1}}

% definition de Couleurs 
% sont prédéfinies : black white red green blue magenta yellow
\definecolor{BleuPhil}{rgb}{0.25,0.25,1}  		% Sur la base deRougeVertBleu
\definecolor{RougePhil}{rgb}{1,0.15,0.15} % Sur la base deRougeVertBleu
\definecolor{VertPhil}{rgb}{0.25,1,0.25}  % Sur la base deRougeVertBleu
\definecolor{MagentaPhil}{rgb}{0.5,0,0.7}  % Sur la base deRougeVertBleu
\definecolor{NoirPhil}{rgb}{0.,0.,0.}  % Sur la base deRougeVertBleu

\def\yaxis#1{\cc{\rotatebox{90}{{\small #1}}}}
\def\th{-th\XS}
\def\mSENSE{\texttt{mSENSE}}
\def\hermit{\mbox{\tiny\textsf{H}}}
\def\bibdebase{revueabr,bibfrabr,biblio,baseTot}
\def\BIBLIO{\bibdebase}

\title{Fast joint detection-estimation of evoked brain activity in event-related \lowercase{f}MRI using a variational approach}

% \graphicspath{{./Figures/}}
\graphicspath{{./Figures/},{./Figures_algorithmic_efficiency/},{./Figures_realdata/}}
\author {Lotfi CHAARI, \textit{Member, IEEE}, Thomas VINCENT, Florence FORBES,\\ Michel DOJAT, \textit{Senior Member, IEEE}
and Philippe CIUCIU, \textit{Senior Member, IEEE}
\thanks{Lotfi CHAARI, Thomas VINCENT and Florence FORBES are with the MISTIS team at INRIA Grenoble Rh\^one-Alpes, 655 avenue de l'Europe, Montbonnot, 38334 Saint
Ismier Cedex, France. Lotfi CHAARI and Thomas VINCENT are also
affiliated to CEA/DSV/$\mathrm{I}^2$BM/Neurospin, CEA Saclay, Bat.
145, Point Courrier 156, 91191 Gif-sur-Yvette cedex, France.
E-mail:
$\{$lotfi.chaari,thomas.vincent,florence.forbes$\}$@inria.fr.
Philippe CIUCIU is with CEA/DSV/$\mathrm{I}^2$BM/Neurospin, CEA
Saclay, Bat. 145, Point Courrier 156, 91191 Gif-sur-Yvette cedex,
France. E-mail: $\{$philippe.ciuciu$\}$@cea.fr. Michel
DOJAT is with INSERM, U836, GIN and University Joseph Fourier,
Grenoble, France. E-mail: Michel.Dojat@ujf-grenoble.fr.} }

% \thanks{A
% preliminary version of this work has been published in
% \cite{Chaari_MICCAI_2011}
\begin{document}
\maketitle
\vspace{-.5cm}
\begin{abstract}

In standard clinical within-subject analyses of event-related fMRI
data, two steps are usually performed separately: detection of
brain activity and estimation of the hemodynamic response. Because
these two steps are inherently linked, we adopt the so-called
region-based Joint Detection-Estimation (JDE) framework that
addresses this joint issue using a multivariate inference for
detection and estimation. JDE is built by making use of a regional
bilinear generative model of the BOLD response and constraining
the parameter estimation by physiological priors using temporal
and spatial information in a Markovian modeling. In contrast to
previous works that use Markov Chain Monte Carlo (MCMC) techniques
to approximate the resulting  intractable posterior distribution,
we recast the JDE into a missing data framework and derive a
Variational Expectation-Maximization (VEM) algorithm for its
inference. A variational approximation is used to approximate the
Markovian model in the unsupervised spatially adaptive JDE
inference, which allows fine automatic tuning of spatial
regularisation parameters. It follows a new algorithm that
exhibits interesting properties compared to the previously used
MCMC-based approach. Experiments on artificial and real data show
that VEM-JDE is robust to model mis-specification and provides
computational gain while maintaining good performance in terms of
activation detection and hemodynamic shape recovery.
% that deals with spatial dependencies between
% voxels belonging to the same functionally homogeneous
% \textit{parcel} in the mask of the 3D brain. After building a
% spatially adaptive General Linear Model, prior information is
% introduced and a hierarchical Bayesian model is established.
\end{abstract}
\vspace{-.5cm}
% \newpage
\section{Introduction}
% These two problems have been addressed sequentially for a long while before the
% Joint Detection Estimation (JDE) framework is proposed.
Functional Magnetic Resonance Imaging~(fMRI) is a powerful tool to
non-invasively study the relationship between a sensory or
cognitive task and the ensuing evoked neural activity through the
neurovascular coupling measured by the BOLD
signal~\cite{Ogawa_90}. Since the 90's, this modality has become
widely used in neuroimaging. Functional connectivity analyses aim
at studying the interactions between signals and thus provide
insight on integrative cerebral phenomena. In a complementary
manner, we focus on the recovery of localization and dynamics of
local evoked activity, thus on specialized cerebral processes. In this
setting, the key issue is the modeling of the link between
stimulation events and the induced BOLD effect throughout the
brain. Physiological non-linear models~\cite{Buxton97,Riera04b_}
are the most specific approaches to properly describe this link
but their computational cost and their identifiability issues
limit their use to a limited number of specific regions and to a
few experimental conditions. The common approach, being the
context of this paper, rather consists of linear systems which
appear more robust and tractable. Here, the link between
stimulation and BOLD effect is modelled through a convolutive
system where each stimulus event induces a BOLD response, via the convolution of the binary stimulus sequence with the 
Hemodynamic Response Function (HRF). It follows two tasks for
such BOLD analysis: the detection of where cerebral activity
occurs and the estimation of its dynamics through the HRF
identification. Commonly, the estimation part is ignored and the
HRF is fixed to a canonical version which has been fitted on
visual areas~\cite{Boynton96}. The detection task is performed by
a General Linear Model (GLM), where stimulus-induced components
are assumed to be known and only their relative weighting are to
be recovered in the form of effect maps~\cite{Friston_94b}.
However, spatial intra-subject and between-subject variability of
the response function has been
highlighted~\cite{Handwerker2004,Badillo11}, in
 addition to potential timing fluctuations induced by the paradigm (eg variations in delay).
To take this variability into account, more flexibility can be injected in the GLM framework by adding more regressors. In a parametric setting,
this amounts to adding a function basis, such as canonical HRF derivatives or a set of gamma functions. In a non-parametric setting, all HRF
 coefficients are explicitely encoded as a Finite Impulse Response~\cite{Henson01}.
The major drawback of these GLM extensions is the multiplicity of regressors for a given condition, so that the detection task is more
difficult
to perform and that statistical power is decreased. Moreover, the more coefficients to recover, the more ill-posed the problem
 becomes.
The alternative approaches which aim at keeping a single regressor
per condition and add also a temporal regularisation constraint to
fix the ill-posedness are the so-called regularized FIR
methods~\cite{Goutte_00,Ciuciu03b}. Still, they do not overcome the
low signal-to-noise ratio inherent to BOLD signals, and they lack
robustness especially in non-activating regions. All the issues
encountered in the previously mentioned approaches are linked to
the sequential treatment of the detection and estimation tasks.
Indeed, these two problems are strongly linked: on the one hand, a
precise localization of brain activated areas strongly depends on
a reliable HRF model;
 on the other hand, a robust estimation of the HRF is only possible in activated areas where enough relevant signal is
measured~\cite{Kershaw99_b}. This consideration led to jointly
perform these two tasks~\cite{Makni05b,dePasquale08} in a Joint
Detection-Estimation (JDE)
framework~\cite{BDSMakni08bis,Vincent10} which is the basis of the
approach developed in this paper. To improve the estimation
robustness, a gain in HRF reproducibility is performed by
spatially aggregating signals so that a constant HRF shape is
locally considered across a small group of voxels, {\it i.e.} a
region or a parcel. The procedure then implies a partitioning of
the data into functionally homogeneous parcels, in the form of a
cerebral parcellation. \rem{ It mainly relies on two key steps:
detecting evoked brain activity and estimating underlying brain
hemodynamics.  Although they are strongly linked, these two
problems have been addressed sequentially for a long while before
the Joint Detection Estimation~(JDE) framework was first proposed
in 2005 \cite{Makni05b}. Processing them jointly allows to enrich
iteratively both steps.
% Proceeding independetly, no possible feedback is possible although the
% two steps are strongly connected.
On the one hand, a precise localization of brain activated areas
 strongly depends on a reliable model for the Finite
Impulse Response (FIR) of the neurovascular coupling also known as
the hemodynamic response function~(HRF).
 On the other hand, a robust estimation of the HRF
is only possible in activated areas where enough relevant signal
is measured. Moreover, to localize which parts of the brain are
activated by a given stimulus, most approaches assume a single
canonical \textit{a priori} model for the HRF~\cite{Friston_94b}.
However, it has been clearly demonstrated that this response can
vary in space and between
subjects~\cite{Handwerker2004,Badillo11}.
% so that
% both issues of properly detecting evoked activity and estimating
% the HRF  play a central role in fMRI data analysis. They are
% usually dealt with independently with no possible feedback
% although they are strongly connected.
The JDE framework  introduced in~\cite{Makni05b,Makni08} and
extended in~\cite{Vincent10} to account for spatial correlation
between neighboring voxels in the brain volume, can also account
for such sources of variability. Since robust and accurate HRF
estimation can only be achieved in regions that elicit an evoked
response to an experimental condition~\cite{Kershaw99_b}, the JDE
approach has been defined at an intermediate spatial resolution
corresponding to \textit{parcels} in which a fair compromise
between homogeneity of the BOLD signal and reproducibility of the
HRF shape is achieved.
}
%Within each parcel, a single HRF shape is estimated while
%simultaneously detecting stimulus-induced activity.
In brief, the JDE approach mainly rests upon: {\em i)} a non-parametric or
FIR modeling of the HRF at this parcel-level for an unconstrained
HRF shape; {\em ii)} prior information about the temporal
smoothness of the HRF to guarantee a physiologically plausible
shape; and {\em iii)} the modeling of spatial correlation between
neighboring voxels within each parcel using condition-specific
discrete hidden Markov fields. In~\cite{Makni08,Vincent10},
posterior inference is carried out in a Bayesian setting using a
Markov Chain Monte Carlo~(MCMC) method, which is computationally
intensive and requires the fine tuning of several parameters.
%Several other attempts to segregate neurological and hemodynamic
%events from fMRI time series have been proposed~(see the
%references in~\cite{Vincent10}).
% Among them lies an interesting bilinear
% dynamical system formulation~\cite{BDSMakni08} that deals
% with unknown HRF and uses Variational Bayes~(VB) approximation for
% tractability.
%%
%% To improve tractability, a recent univariate approach has been
%% proposed~\cite{BDSMakni08} based on Variational Bayes~(VB)
%% approximation. However, this approach remains univariate and does
%% not consider spatial correlation between voxels.
%%
%  that does not consider spatial correlation between voxels.
% However, from a spatial viewpoint, this work remains
% univariate and ignores spatial correlation between voxels.

In this paper, we reformulate the complete JDE
framework~\cite{Vincent10} as a mis\-sing data problem and propose
a simplification of its estimation procedure.
% Akin to~\cite{BDSMakni08},
We resort to a variational approximation using a Variational
Expectation Maximization~(VEM) algorithm in order to derive
estimates of the HRF and stimulus-related activity. Variational
approximations have been widely and successfully employed in the
context of fMRI analysis: to model auto-regressive noise in the
context of a Bayesian GLM~\cite{Penny03_b}, to characterize
cerebral hierarchical dynamic models~\cite{Friston07_}, to model
transient neuronal signals in a Bayesian dynamical
system~\cite{BDSMakni08bis}, or to perform inference of spatial
mixture models for the segmentation of GLM effects
maps~\cite{Woolrich06}. As in our study, the primary objective of
resorting to variational approximations is to alleviate the
computational burden associated with stochastic MCMC approaches.
Akin to~\cite{Woolrich06}, we aim at comparing the stochastic and
variational based inference schemes but on the more complex matter
of detecting activation \emph{and} estimating the HRF
whereas~\cite{Woolrich06} treated only a detection (or
segmentation) problem.

Compared to a JDE MCMC implementation, the proposed approach does not require priors on the model parameters for inference to be
carried out. However, for more robustness and to
make the proposed approach completely auto-calibrated, the adopted
model may be extended by injecting additional priors on some of its parameters.

Experiments on artificial and real data demonstrate the good
performance of our VEM algorithm. Compared to the MCMC
implementation, VEM is more computationally efficient, is robust
to mis-specification of the parameters, to deviations from the
model and adaptable to various experimental conditions.
This  increases considerably the potential impact of the JDE
framework and makes its application to fMRI studies in
neuroscience easier and more valuable. This new  framework has
also the advantage of providing straightforward criteria for model
selection.

The rest of this paper is organized as follows. In
Section~\ref{JDEsec}, we introduce the hierarchical Bayesian model
for the JDE
 framework in the within-subject fMRI context.
In Section~\ref{sec:VEM},  the VEM algorithm based on variational
approximations  for inference is described. Evaluation on real and
artificial fMRI datasets are reported in Section~\ref{simuls} and
the performance comparison between the MCMC and VEM
implementations is reported in Section~\ref{sec_algorithmics_efficiency}. Finally,
Section~\ref{sec:conclusion} concludes with a discussion of the
points highlighted by the approach and areas for further research.

\section{Bayesian framework for the joint detection-estimation}\label{JDEsec}

Matrices and vectors are denoted with bold upper and lower case
letters~({\it e.g.} \Pb and $\mub$). A vector is by convention a
column vector. The transpose is denoted by $\T$. Unless stated
otherwise, subscripts $j$, $m$, $i$ and $n$ are respectively
indexes over voxels, stimulus types, mixture components and time
point. The Gaussian distribution with mean $\mub$ and covariance
matrix  $\Sigmab$ is denoted by $\Nc(\mub, \Sigmab)$.

\subsection{The parcel-based model}

 We first recast the parcel-based JDE model of~\cite{Makni08,Vincent10} in a missing data framework.
Let us assume that the brain is decomposed in
$\Pcb=(\Pc_\gamma)_{\gamma=1:\Gamma}$ parcels, each of them having
homogeneous hemodynamic properties. The fMRI time series $\yb_j$
is measured in voxel $j \in \Pc_\gamma$ at times $(t_n)_{n=1
\ldots N}$, where $t_n=n\TR$, $N$ being the number of scans and
$\TR$, the time of repetition. The number of different stimulus
types or experimental conditions is $M$.  For a given parcel
$\Pc_\gamma$ containing a group of connected voxels, a unique BOLD
signal model is used in order to link the observed data
$\Yb=\{\yb_j\in\RR^N, j \in \Pc_\gamma\}$ to the HRF $\hb_\gamma
\in \RR^{D+1}$ specific to $\Pc_\gamma$ and to the response
amplitudes $\Ab = \{\ab^m, m=1\ldots M\}$ with $\ab^m=\{a^m_j,
j\in \Pc_\gamma\}$ and $a_j^m$ being the magnitude at voxel $j$
for condition $m$. More specifically, the observation model at
each voxel $j\in\Pc_\gamma$ is expressed as follows
\cite{Makni08}:
\begin{align}%\label{ObservationModel}
\label{eq_dec_modele_generatif_lti}
\yb_j &= \Sb_j\hb_\gamma + \Pb \ellb_j + \varepsilonb_j,\quad \text{with} \quad
\Sb_j= \sum_{m=1}^M a_j^m \Xb_m
\end{align}
where $\Sb_j\hb_\gamma$ is the summation of the stimulus-induced
components of the BOLD signal. The binary matrix
$\Xb_m=\{x^{n-d\Delta t}_m, n=1\ldots N, d=0 \ldots D\}$ is of
size $N\times (D+1)$ and provides information on the stimulus
occurrences for the $m$\th experimental condition, $\Delta t < TR$
being the sampling period of the unknown HRF $\hb_\gamma=
\{h_{d\Delta t}, d=0 \ldots D\}$ in $\Pc_\gamma$. This hemodynamic
response is a consequence of the neuronal excitation which is
commonly assumed to occur following stimulation. The scalars
$a_j^m$'s are weights that model the transition between
stimulations whose occurrences are informed by the $\Xb_m$
matrices ($m=0 \ldots M$), and the vascular response informed by
the filter $\hb_\gamma$. It follows that the $a_j^m$'s are
generally referred to as Neural Response Levels (NRL). The rest of
the signal is made of
 matrix $\Pb$, which corresponds to physiological artifacts accounted for via a  low frequency
 orthonormal function basis of size $N\times O$. At each voxel $j$ is associated a vector of low
frequency drifts $\ellb_j \in \RR^O$ which has to be estimated.
Within parcel $\Pc_\gamma$, these vectors may be grouped into the
same matrix $\Lb = \{\ellb_j, j \in \Pc_\gamma\}$. Regarding the
observation noise, the $\varepsilonb_j$'s are  assumed to be
independent with $\varepsilonb_j \sim \Nc(0, \Gammab_j\M)$ at
voxel~$j$ (see Section~\ref{sec:likelihood} for more details). The
set of all unknown precision matrices (inverse of the covariance
matrices) is denoted by $\Gammab=\acc{\Gammab_j, j \in
\Pc_\gamma}$.

Finally, detection is handled through the introduction of
activation class assignments \linebreak $\Qb=\bigacc{\qb^m, m=1\ldots
M}$ where $\qb^m = \bigacc{q_j^m, j \in \Pc_\gamma}$ and $q_j^m$
represents the {\it activation class} at voxel $j$ for
experimental condition $m$. The NRL coefficients will therefore be
expressed conditionally to these hidden variables. In other words,
the NRL coefficients will depend on the activation status of the
voxel $j$, which itself depends on the activation status of
neighbouring voxels thanks to a Markov model used as a spatial
prior on $\Qb$ (\textit{cf} Section~\ref{par:Q_prior}). Without
loss of generality, here the number of classes is $I=2$ for
activated and non-activated voxels. An additional deactivation
class ($I=3$) may be considered depending on the experiment. In
the following developments, all provided formulas are general
enough to cover this case.

\subsection{A hierarchical Bayesian Model}
In a Bayesian framework, we first need to define the likelihood
and prior distributions for the model variables
$(\Yb,\Ab,\hb_\gamma,\Qb)$ and parameters $(\Thetab)$. Using the
hierarchical structure between $\Yb$, $\Ab$, $\hb_\gamma$, $\Qb$
and $\Thetab$, the complete model is  given by the joint
distribution of both the observed and unobserved (or missing) data: $p(\Yb,
\Ab, \hb_\gamma, \Qb;\Thetab) = p(\Yb \I \Ab, \hb_\gamma;\Thetab)
\; \;p(\hb_\gamma;\Thetab) p(\Ab \I \Qb;\Thetab)
\;p(\Qb;\Thetab).$ To fully define the hierarchical model, we now
specify each term in turn.
% \noindent {\bf The $p(y \I a, h)$ term.}
\subsubsection{Likelihood}\hfill \\
\label{sec:likelihood} The definition of the likelihood depends on
the noise model. In \cite{Woolrich01_b,Makni_06}, an
autoregressive (AR) noise model has been adopted to account for
serial correlations in fMRI time series. It has also been shown in
\cite{Makni_06} that a spatially-varying first-order AR noise
model helped controlling false positive rate. In the same context,
we will assume such a noise model
$\varepsilonb_j\sim\Nc(\zerob,\Gammab_j\M)$ with $\Gammab_j
=\sigma_j^{-2}\Lambdab_j$ where $\Lambdab_j$ is a tridiagonal
symmetric matrix which depends on the AR(1) parameter
$\rho_j$~\cite{Makni08}: $(\Lambdab_j)_{1,1}=(\Lambdab_j)_{N,N} =
1$,  $(\Lambdab_j)_{n,n} = 1 + \rho_j^2 $ for $n=2:N-1$ and
$(\Lambdab_j)_{n+1,n}=(\Lambdab_j)_{n,n+1}=-\rho_j$ for $n=1:N-1$.
These parameters are assumed voxel-varying due to their
tissue-dependence~\cite{Woolrich04_b,Penny07_b}. The likelihood
can therefore be decomposed as:
\begin{align}
\label{eq:dec_vraisemblance}
p(\Yb\I\Ab, \hb_\gamma;\Lb,\Gammab) %&= \prod_{j=1}^J p(\yb_j\I\hb,\ab_j,\ellb_j,\theta_{0,j}) \nonumber\\
&\propto\prod_{j \in \Pc_\gamma} |\Gammab_j|^{\froc{-1}2}
\exp\bigpth{-\frac12\widebar{\yb}_j\T\Gammab_j\widebar{\yb}_j},
\end{align}
where $|\Gammab_j| = \sigma_j^{-2N}\,|\Lambdab_j|$, $|\Lambdab_j|=
1-\rho_j^2$ and $\widebar{\yb}_j = \yb_j -\Pb \ellb_j -\Sb_j
\hb_\gamma$.

% From~\eqref{eq_dec_modele_generatif_lti}, it comes that  $p(y \I a, h) =
% \prod\limits_{i\in \Pc} p(y_i \I a_i, h)$ with
% $Y_i\I (A_i=a_i,H=h) \sim
% \Nc\Bigpth{\sum_{m=1}^M a_{mi} \Xb_m h + \Pb\ell_i ,\Gammab\M_i}$.\\
\subsubsection{Model priors}
\paragraph{HRF}\hfill \\
Akin to~\cite{Makni08,Vincent10}, we introduce constraints in the
HRF prior that favor smooth variations in $\hb_\gamma$ by
controlling its second order derivative: $ \hb_\gamma \sim \Nc(0,
v_\hb\Rb) \; \WITH  \Rb = (\Delta t)^4 \;(\Db_2\T \Db_2)\M$ where
$\Db_2$ is the second-order finite difference matrix and $v_\hb$
is a parameter to be estimated. Moreover, boundary constraints
have also been fixed on $\hb_\gamma$ as
in~\cite{Makni08,Vincent10} so that $h_0 = h_{D \Delta t}=0$.

\paragraph{Neuronal response levels}\hfill \\
Akin to~\cite{Makni08,Vincent10}, the NRLs $a_{j}^m$ are assumed
to be statistically independent across conditions:
$p(\Ab;\thetaA)=\prod\limits_m p(\ab^m;\thetab_m)$ where
  $\thetaA=\{\thetab_m, m=1 \ldots M\}$  and $\thetab_m$ gathers the parameters for the $m$\th condition.
 A mixture model is then adopted by using the assignment
variables $q_j^m$ to segregate non-activated voxels ($q_j^m=1$) from
activated ones ($q_j^m=2$). For the $m$\th condition, and
conditionally to the assignment variables $\qb^m$, the NRLs are
assumed to be independent: $p(\ab^m\I
\qb^m;\thetab_m)=\prod\limits_{j\in \Pc_\gamma} p(a_j^m\I
q_j^m;\thetab_m)$. If $q_j^m=i$ then $p(a_j^m\I q_j^m=i;\thetab_m)
\sim \Nc(\mu_{im},v_{im})$.
% where $\Nc(\; . \;  ; \mu_{im},v_{im})$ denotes the Gaussian
%distribution with mean $\mu_{im}$ and variance $v_{im}$.
 The Gaussian
parameters $\thetab_m=\{\mu_{im},v_{im}, i=1 \ldots I \}$ are
unknown. We denote by $\mub\!=\!\!\{ \mub_m, m=1 \ldots M \}$ with
$\mub_m = \{\mu_{1m},\ldots,\mu_{Im}\}$ and $\vb= \{\vb_m, m=1
\ldots M\}$ with $\vb_m=\{v_{1m},\ldots, v_{Im}\}$. More
specifically, for non-activating voxels we set for all $m$,
$\mu_{1m}\!=\!0$.
% (without loss of generality, only two classes are considered here: $I=1$)

\paragraph{Activation classes}\hfill \\
\label{par:Q_prior}
% \noindent {\bf The $p(z)$ term.}
As in~\cite{Vincent10}, we assume prior independence between the
$M$ experimental conditions regarding the activation class
assignments. It follows that  $p(\Qb) \!=\! \prod\limits_{m=1}^M
p(\qb^m ; \beta_m)$  where we assume in addition that $p(\qb^m;
\beta_m)$ is a spatial Markov prior, namely a Potts model with
interaction parameter $\beta_m$ \cite{Vincent10}:
\begin{align}
\label{IsingMRF} p(\qb^m;\beta_m) &= Z(\beta_m)\M
\exp\bigpth{\beta_m U(\qb^m) }\quad \text{with} \quad U(\qb^m) =
\sum_{{j \thicksim k}} \scalarprod{q_j^m}{q_k^m}
\end{align}
and where $Z(\beta_m)$ is the normalizing constant and for all
$(a,b) \in \RR^2\;,\;\scalarprod{a}{b}=1$ if $a=b$ and 0
otherwise. The notation $j\thicksim k$ means that the summation is
over all neighboring voxels. Moreover, the neighboring system may
cover a 3D scheme through the brain volume. The unknown parameters
are denoted by
$\betab= \acc{\beta_m, m=1 \ldots M}$. In what follows, we will consider a 6-connexity 3D neighboring system.

 For the complete model, the whole
set of parameters is denoted by $\Thetab = \acc{\Gammab, \Lb,
\mub, \vb, \vb_h, \betab}$ and belong to  a set
$\underline{\Thetab}$.

\section{Estimation by variational Expectation-Maximization}\label{sec:VEM}
% \vspace{0.1cm}
We propose to use an Expectation-Maximization~(EM) framework to
deal with the missing data namely, $\Ab \in \Ae$, $\hb_\gamma \in \He$, $\Qb
\in \Qe$.
 Let $\DQe$ be the set of all probability distributions on
$\Ae \times \He \times \Qe$. EM can be viewed~\cite{Neal98} as an
alternating maximization procedure of a function $\Fc$ on $\DQe$,
$\Fc(p, \Thetab) = \Espud{p}{\log p(\Yb, \Ab, \hb_\gamma, \Qb \I
\Thetab)} + \Gc(p)$ where $\Espud{p}{.}$ denotes the expectation
with respect to $p$ and $\Gc(p) = - \Espud{p}{\log p(\Ab,
\hb_\gamma, \Qb)}$ is the entropy of $p$. At iteration $(r)$,
denoting the current parameter values by $\Thetab^{(r-1)}$, the
alternating procedure proceeds as follows:

\vspace{-0.2in}
\begin{align}
  \label{eq:Step1GAM}
  \text{{\bf E-step:} }& p^{(r)}_{A,H_\gamma,Q}
  =\argmax\limits_{p \in \DQe} \; \Fc(p, \Thetab^{(r-1)})\\
  \text{{\bf M-step:} }& \Thetab^{(r)}  = \argmax\limits_{ \Theta \in \underline{\Thetab}} \; \Fc(p^{(r)}_{A,H_\gamma,Q}, \Thetab)
  \label{eq:Step2GAM}
\end{align}

The optimization step in Eq.~\eqref{eq:Step1GAM} leads to
$p^{(r)}_{A,H_\gamma,Q} = p(\Ab,\hb_\gamma,\Qb \I \Yb ,
\Thetab^{(r-1)})$, which is intractable for our model. Hence, we
resort to a variational EM variant in which the intractable
posterior is approximated as a product of three pdfs on $\Ae$,
$\He$ and $\Qe$ respectively.

\vspace{.3cm}
%\subsection{Variational EM algorithm (VEM)}
Previous attempts to use variational  inference \cite{Beal03b_} in
fMRI~\cite{Woolrich06,Penny03_b} have been successful with this
type of approximations usually validated by assessing its fidelity
to its MCMC counterpart. In Section~\ref{simuls}, we will also
provide such a comparison. The fact that the HRF $\hb_\gamma$ can
be equivalently considered as missing variables or random
parameters induces some similarity between our Variational EM
variant and the Variational Bayesian EM
 algorithm presented in \cite{Beal03b_}. Our
framework varies slightly from the case of conjugate exponential
models described in \cite{Beal03b_} and more importantly, our
 presentation offers the possibility to deal with
extra parameters $\Thetab$ for which prior information may not
be available. \\
We propose here to use an EM variant in which the intractable
E-step is instead solved over $\tilde{\Dc}$, a restricted class of
probability distributions chosen as the set of distributions that
factorize as $\wt{p}_{A,H_\gamma,Q} = \wt{p}_A \wt{p}_{H_\gamma}
\wt{p}_Q$ where $\wt{p}_A$, $\wt{p}_{H_\gamma}$ and $\wt{p}_Q$ are
probability distributions on $\Ae$, $\He$ and $\Qe$, respectively.
It follows then that our E-step becomes an approximate E-step,
which can be further decomposed into three stages that consist of
updating the three pdfs,   $\wt{p}_{H_\gamma}$, $\wt{p}_A$  and
 $\wt{p}_Q$,  in turn
using three equivalent expressions of $\Fc$ when $p$ factorizes as
in $\tilde{\DQe}$. At iteration $(r)$ with current estimates
denoted by $\wt{p}_H^{(r-1)}, \wt{p}_A^{(r-1)}, \wt{p}_Q^{(r-1)}$  and
$\Thetab^{(r-1)}$,
 the updating rules become:
\begin{align}
\text{{\bf E-H:} }\; \wt{p}_{H_\gamma}^{(r)} &=
\argmax_{p_{H_\gamma} } \Fc(\wt{p}_{A}^{(r-1)}\; p_{H_\gamma}\;%\in \Dc_H
\wt{p}_Q^{(r-1)} ,\Thetab^{(r-1)}) \nonumber \\
\text{{\bf E-A:} } \; \wt{p}_A^{(r)} &= \argmax_{p_A}%\in\Dc_A
 \Fc(p_A \; \wt{p}_{H_\gamma}^{(r)} \; \wt{p}_Q^{(r-1)} , \Thetab^{(r-1)})\nonumber \\
 \text{{\bf E-Q:} } \; \wt{p}_Q^{(r)} &= \argmax_{p_Q } %\in \Dc_Q
\Fc(\wt{p}_A^{(r)} \; \wt{p}_{H_\gamma}^{(r)} \; p_Q,
\Thetab^{(r-1)}).\nonumber
\end{align}

\noindent Introducing the Kullback-Leibler divergence between
$\wt{p}_{A,H_\gamma,Q}$ and $p_{A,H_\gamma,Q}$\;, we have
\begin{align}\label{eq:defKL}
\Dc(\wt{p}_{A,H_\gamma,Q}\,||\, p_{A,H_\gamma,Q} )&=\int \wt{p}_{A,H_\gamma,Q}(\Ab,\hb_\gamma,  \Qb)
\log \frac{\wt{p}_{A,H_\gamma,Q}(\Ab,\hb_\gamma,  \Qb)}{p_{A,H_\gamma,Q}(\Ab,\hb_\gamma,  \Qb)}
\dd\Ab\dd \hb_\gamma  \dd\Qb.
\end{align}
According to~\cite{Neal98}, we also have $\Fc(\wt{p}_{A,H_\gamma,Q}, \Thetab) = \log
p(\Yb;\Thetab)- \Dc(\wt{p}_{A,H_\gamma,Q}\,||\,
p_{A,H_\gamma,Q})$ so that the steps above can be
equivalently written in terms of minimizations of the
Kullback-Leibler divergence. The properties of the latter lead to
the following solutions:
\begin{align}
\text{{\bf E-H:} }\;   \wt{p}_{H_\gamma}^{(r)}(\hb_\gamma)
&\propto \exp\pth{\Espud{\wt{p}_{A}^{(r-1)} \wt{p}_Q^{(r-1)}}{\log
p(\hb_\gamma \I \Yb, \Ab, \Qb;
\Thetab^{(r-1)}}} \label{eq:E-H}\\
\text{{\bf E-A:} } \; \wt{p}_A^{(r)}(\Ab) &\propto \exp\pth{
\Espud{\wt{p}_{H_\gamma}^{(r)} \wt{p}_Q^{(r-1)}}{\log p(\Ab \I
\Yb, \hb_\gamma, \Qb; \Thetab^{(r-1)})}
}\label{eq:E-A}\\
\text{{\bf E-Q:} } \; \wt{p}_Q^{(r)}(\Qb) &\propto \exp\pth{
\Espud{\wt{p}_A^{(r)} \wt{p}_{H_\gamma}^{(r)} }{\log p(\Qb \I \Yb,
\Ab, \hb_\gamma ; \Thetab^{(r-1)}) }} \; \label{eq:E-Q}.
\end{align}
   \noindent The corresponding {\bf M-step} is (since $\Thetab$ and $\Gc(p^{(r)}_{A, H_\gamma, Q})$ are independent):
\begin{align}
 % \label{eq:StepM}
 \qquad \text{{\bf M:} }\; \Thetab^{(r)}  &=
\argmax_{\Thetab} \;  \Espud{\wt{p}_A^{(r)}
\wt{p}_{H_\gamma}^{(r)}\wt{p}_Q^{(r)}}{\log p(\Yb, \Ab, \hb_\gamma, \Qb ; \Thetab)} \;\label{eq:M}.
\end{align}
These steps lead to explicit calculations for
$\wt{p}^{(r)}_{H_\gamma}$, $\wt{p}^{(r)}_A$, $\wt{p}^{(r)}_Q$ and
the parameter set \linebreak $\Thetab^{(r)}=\acc{\Gammab^{(r)},
\Lb^{(r)},\mub^{(r)}, \vb^{(r)}, \vb_h^{(r)}, \betab^{(r)}}$.

\begin{itemize}
 \item[\textbullet] \text{{\bf E-H} step: } From Eq.~\eqref{eq:E-H} standard algebra enables to derive that $\wt{p}^{(r)}_{H_\gamma}$ is a Gaussian
distribution  $\wt{p}^{(r)}_{H_\gamma} \sim
\Nc(\mb_{H_\gamma}^{(r)}, \Sigmab_{H_\gamma}^{(r)})$ whose
parameters are detailed in  Appendix~\ref{append:E-H}.  The
expressions for $\mb_{H_\gamma}^{(r)}$ and
$\Sigmab_{H_\gamma}^{(r)}$ are similar to those derived in the
MCMC case~\cite[Eq.~(B.1)]{Makni08} with expressions involving the
$a_{j}^m$'s replaced by their expectations wrt
$\wt{p}_{A}^{(r-1)}$.

 \item[\textbullet] \text{{\bf E-A} step :}  Using Eq.~\eqref{eq:E-A},
standard algebra rules allow to identify the Gaussian distribution of
$\wt{p}^{(r)}_A$ which writes as
$\wt{p}^{(r)}_A = \prod_{j \in \Pc_\gamma} \wt{p}^{(r)}_{A_j}$
with $\wt{p}^{(r)}_{A_j} \sim \Nc(\mb_{A_j}^{(r)}, \Sigmab_{A_j}^{(r)})$.
More detail about the update of $\wt{p}^{(r)}_A$ is given in Appendix~\ref{append:E-A}.
The relationship with the MCMC update of $\Ab$
is not straightforward. In \cite{Makni08,Vincent10}, the
$a_{j}^m$'s are sampled
 independently and conditionally on the $q_{j}^m$'s. This is not the case in the VEM framework
but some similarity appears if  we set the probabilities
$\wt{p}^{(r-1)}_{Q_{j}^m}(i)$  either to 0 or 1 and consider only
the diago\-nal part of  $\Sigmab_{A_j}^{(r)}$.
 \item[\textbullet] \text{{\bf E-Q} step: }  Using the expressions of $p(\Ab | \Qb)$ and $p(\Qb)$ in Section~\ref{JDEsec},
 Eq.~\eqref{eq:E-Q} yields $\wt{p}^{(r)}_Q(\Qb)\!=\!\!\prod\limits_{m=1}^M\wt{p}^{(r)}_{Q^m}(\qb^m)$
which is intractable due to the Markov prior. To overcome this
difficulty, a number of approximation techniques are available.
To decrease the computational complexity of our EM
algorithm or to avoid introducing additional variables as done
in~\cite{Woolrich06}, we use a mean-field like algorithm which
consists of fixing the neighbours to their mean value.
Following~\cite{Celeux03}, $\wt{p}_{Q^m}^{(r)}(\qb^m)$ can be
approximated by a factorized density
 $\wt{p}_{Q^m}^{(r)}(\qb^m) = \prod\limits_{j \in \Pc_\gamma}
\wt{p}_{Q_{j}^m}^{(r)}(q_{j}^m)$ such that if $q_{j}^m=i$,
\begin{equation}
\wt{p}_{Q_{j}^m}^{(r)}(i)  \propto \Nc(\mb_{A_j^m}^{(r)} ;
\mu_{im}^{(r-1)} , v_{im}^{(r-1)})  f(q_{j}^m=i \I \tilde{q}_{\sim
j}^m; \beta_m^{(r-1)}, \vb_{m}^{(r-1)}),
\end{equation}
where $\tilde{\qb}^m$ is a particular configuration of $\qb^m$
updated at each iteration according to a specific scheme, $\sim j$
denotes neighbouring voxels to $j$ on the brain volume and
$f(q_{j}^m=i \I \tilde{q}_{\sim j}^m; \beta_m^{(r-1)},
\vb_{m}^{(r-1)}) \propto \exp\{\frac{v_{A_j^m
A_j^m}^{(r)}}{v_{im}^{(r-1)}} +\beta_m^{(r-1)} \sum\limits_{k \sim
j} \scalarprod{\tilde{q}_{k}^m}{i}\}$. Hereabove, $m_{A_{j}^m
}^{(r)}$ and $v_{A_{j}^m A_{j}^{m'}}^{(r)}$ denote the $m$  and
$(m,m')$  entries of the mean vector ($\mb_{A_j}^{(r)}$) and
covariance matrix ($\Sigmab_{A_j}^{(r)}$), respectively. The
Gaussian distribution with mean $\mu_{im}$ and variance $v_{im}$
is denoted by $\Nc(\; . \;  ; \mu_{im},v_{im})$, while
$\tilde{q}_{\sim j}^m =\{\tilde{q}_{k}^m, k \sim j\}$. More
details are given in Appendix~\ref{append:E-Q}.

% $\Nc(\; . \;  ;
% \mu_{im},v_{im})$ denotes the Gaussian distribution with mean
% $\mu_{im}$ and variance $v_{im}$ and $\tilde{q}_{\sim j}^m
% =\{\tilde{q}_{k}^m, k \sim j\}$. More detail is given in
% Appendix~\ref{append:E-Q}.

%\begin{equation}
%\label{eq:post-ising}
%\wt{p}^{(r)}_{Q^m}(\qb^m) \!=\! f(\qb^m |
%\ab^m\!=\!\mb_{A^m}^{(r)} ;
%\mub_{m}^{(r-1)},\vb_{m}^{(r-1)},\beta_{m}^{(r-1)}).
%\end{equation}

\item[\textbullet] \text{{\bf M }step: } For this maximization step, we can first rewrite Eq.~\eqref{eq:M} as
\begin{align}
 \label{eq:M-4}
\Thetab^{(r)}  &= \argmax_{\Thetab} \;  \Big[
\Espud{\wt{p}_A^{(r)} \wt{p}_{H_\gamma}^{(r)}}{\log p(\Yb \I \Ab,
\hb_\gamma ; \Lb, \Gammab)} +
\Espud{\wt{p}_A^{(r)} \wt{p}_Q^{(r)}}{\log p(\Ab\I \Qb ; \mub, \vb)}\nonumber \\
 &+
\Espud{\wt{p}_{H_\gamma}^{(r)}}{\log p(\hb_\gamma; \vb_h)} +
\Espud{\wt{p}_Q^{(r)}}{\log p(\Qb ; \betab)}
\Big].
\end{align}

The M-step can therefore be decoupled into four sub-steps
involving separately $(\Lb, \Gammab)$, $(\mub, \vb)$, $\vb_h$ and $\betab$. Some of these sub-steps admit closed-form expressions,
while some other require resorting to iterative or alternate optimization. For more details about related calculations, the interested reader
can refer to Appendix~\ref{append:M}.

\end{itemize}

\section{Validation of the proposed approach}\label{simuls}

This section aims at validating the proposed variational approach.  Simulated
and real contexts  are considered respectively in
sub-sections~\ref{aubsec:simulated}
 and \ref{aubsec:real}. To corroborate the effectiveness of the proposed method, comparisons with its
MCMC counterpart will also be conducted throughout the present
section.

\subsection{Artificial fMRI datasets}
\label{aubsec:simulated}
\begin{figure}[!ht]
\centering
\begin{tabular}{c c c c c}
\hspace{-0.4cm} \small  $\Sb_j$& \hspace{-0.4cm} \small  $\hb_\gamma$& \hspace{-0.4cm}  \small  $\varepsilon_j$ +  $\Pb \ell_j$ &\hspace{-0.4cm} \small  $\yb_j$\\
\includegraphics[height=2.5cm,width=3.5cm]{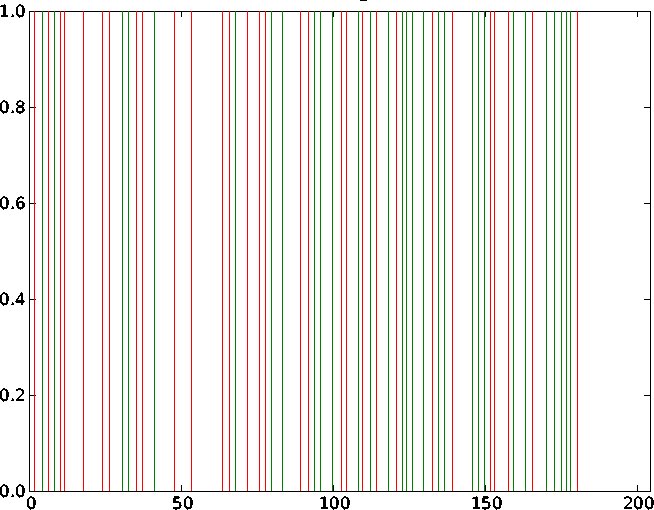} \raisebox{0.7cm}{$\star$}&
\hspace{-0.4cm} \includegraphics[height=2.5cm,width=3.5cm]{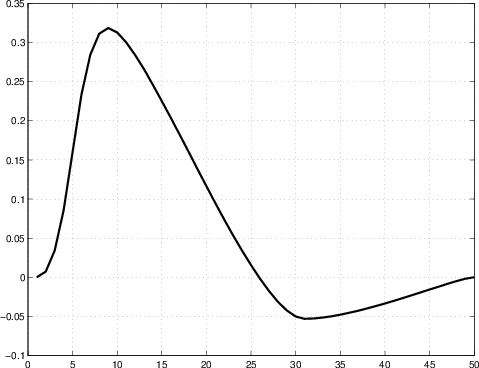}\raisebox{0.7cm}{$+$}&
\hspace{-0.5cm} \includegraphics[height=2.5cm,width=3.5cm]{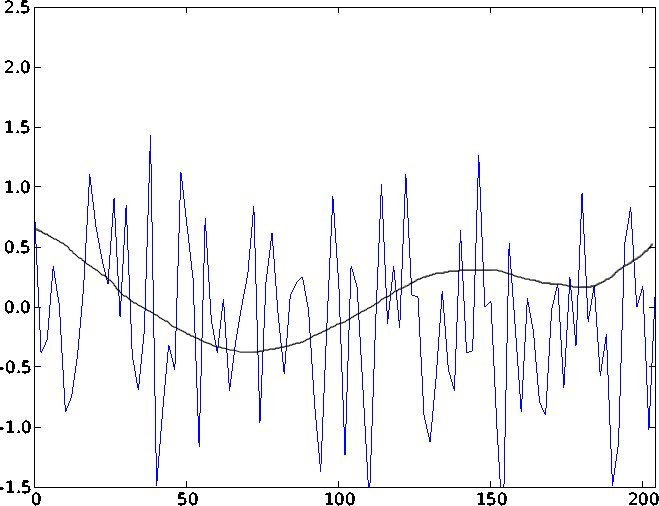} \raisebox{0.7cm}{=}& \hspace{-0.4cm}
\hspace{-0.3cm} \includegraphics[height=2.5cm,width=3.5cm]{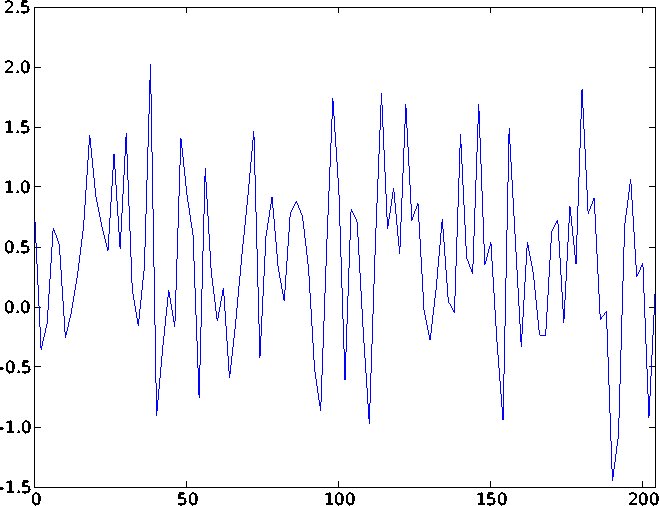}\\
\hspace{-0.4cm} \small scan number &\hspace{-0.4cm} \small Time in s &  \hspace{-0.4cm} \small scan number & \hspace{-0.4cm} \small scan number
\end{tabular}\vspace{-0.3cm}
\caption{\small From left to right: $\Sb_j$ represents the experimental
conditions for voxel $j$ as the scan number increases. They are
convoluted with the hemodynamic model $\hb_\gamma$  of parcel
$\Pc_\gamma$; $\varepsilon_j$ +  $\Pb \ell_j$ represents the noise
and artifact components for voxel $j$ with respect to the
 scan number and $\yb_j$ is the final fMRI time series simulated at voxel
$j$.\label{SyntheticData}}
\end{figure}

In this section, experiments have been conducted on data simulated according to the observation model in
Eq.~\eqref{eq_dec_modele_generatif_lti} where \Pb has been defined from a cosine transform basis as in \cite{Makni08}.
The simulation process is illustrated in Fig~\ref{SyntheticData}.

Different studies have then been conducted in order to validate the detection-estimation performance and robustness. For each of these
studies, some simulation parameters have been changed such as the noise or the paradigm properties. Changing these parameters
aims at providing for each simulation context a realistic BOLD signal.

\subsubsection{Detection-Estimation performance}
\begin{figure}[!ht]
\centering
\begin{tabular}{c c c c c}
&  Ground Truth&  MCMC& VEM&\\
\raisebox{1.5cm}{ $m=1$}&
\includegraphics[height=3cm,width=3cm]{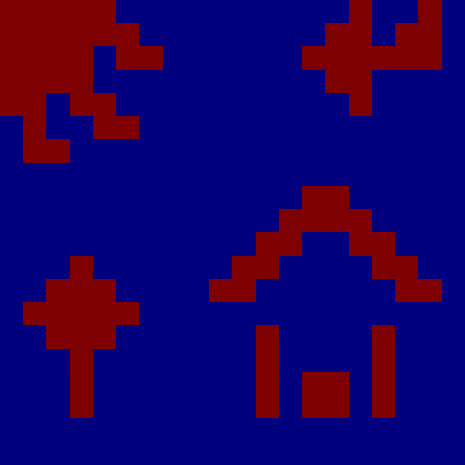}&
\includegraphics[height=3cm,width=3cm]{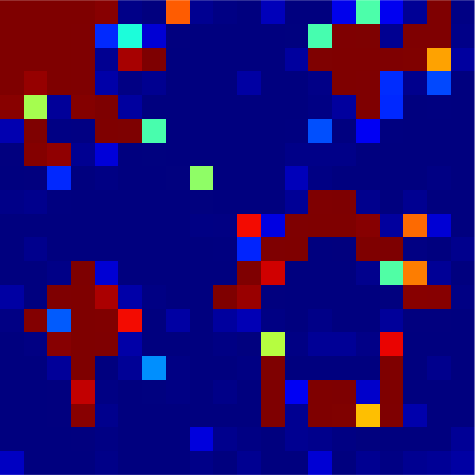}&
\includegraphics[height=3cm,width=3cm]{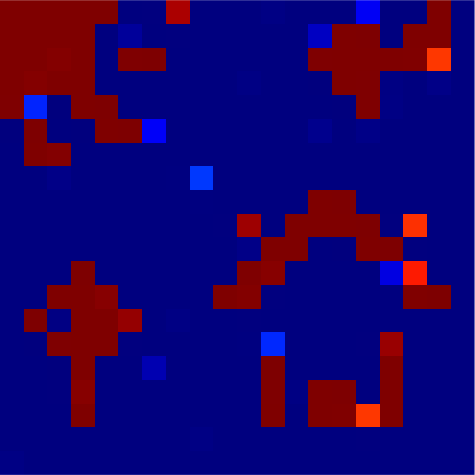}&
\includegraphics[height=3cm,width=.6cm]{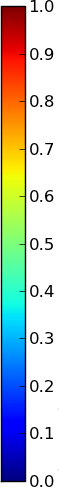}\\
\raisebox{1.5cm}{ $m=2$}&
\includegraphics[height=3cm,width=3cm]{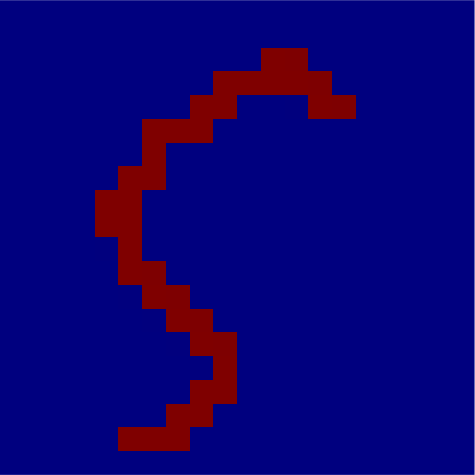}&
\includegraphics[height=3cm,width=3cm]{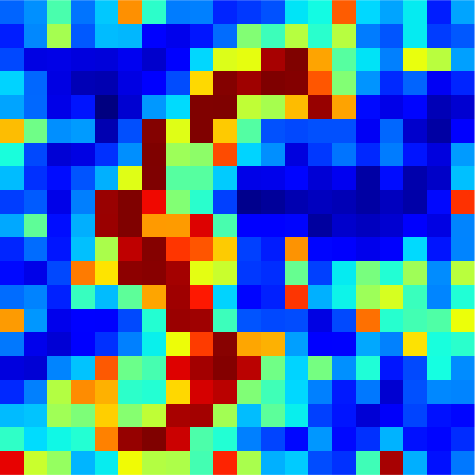}&
\includegraphics[height=3cm,width=3cm]{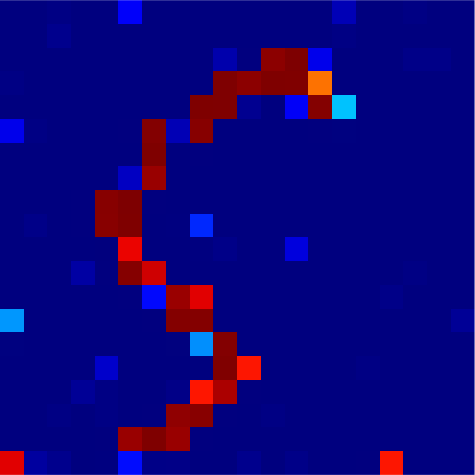}&
\includegraphics[height=3cm,width=.6cm]{colorbar_labels.png}
\end{tabular}
\vspace{-.3cm}
\caption{ Ground truth ({\bf left}) and estimated Posterior
probability maps~(PPM) using MCMC ({\bf middle}) and VEM~({\bf
right}). Condition $m=2$ (Bottom row) represents a lower CNR than condition
$m=1$ (top row). \label{fig:labels}}
\end{figure}

\hfill \\
\label{sec:perf} Data have been simulated here with a Gaussian white
noise $\Gammab\M_j = 1.2~\Ib_N$ ($\Ib_N$ is the $N \times
N$ identity matrix). Two experimental conditions have also been
considered ($M=2$) while ensuring stimulus-varying
Contrast-to-Noise Ratios (CNR)\footnote{For two Regions of Interest (ROI),
${\rm CNR} = 2(\mu_1 - \mu_2)^2/(v_1 + v_2)$ where $(\mu_1,v_1)$ (resp. $(\mu_2,v_2)$) are the intensity mean and variance
within ROI 1 (resp. ROI 2).} achieved by setting
$\mu_{12}=2.8,v_{12}=0.5^2$ and $\mu_{22}=1.8,v_{22}=0.5^2$ so
that higher CNR is simulated for the first experimental condition ($m=1$) compared to the second one ($m=2$).
% $\mu_{12}/v_{12}>\mu_{22}/v_{22}$ (the CNR for $m=1$ is 1.5 times the CNR for  $m=2$).
For each of these conditions, the {\it
initial} artificial paradigm comprised 30 stimulus events. The
simulation process finally yielded time-series of 268 time-points.
Condition-specific activating and non-activating voxels were
defined as 20$\times$20 2D slices as shown in
Fig.~\ref{fig:labels}[left], respectively.

\begin{figure}[!ht]
\centering
\begin{tabular}{c c c c}
&$m=1$ & &$m=2$\\
\rotatebox{90}{ \hspace{1.1cm} \small sensitivity}&
\includegraphics[height=4cm, width=6cm, trim= 1.6cm 1cm 1.5cm 1.3cm, clip=true]{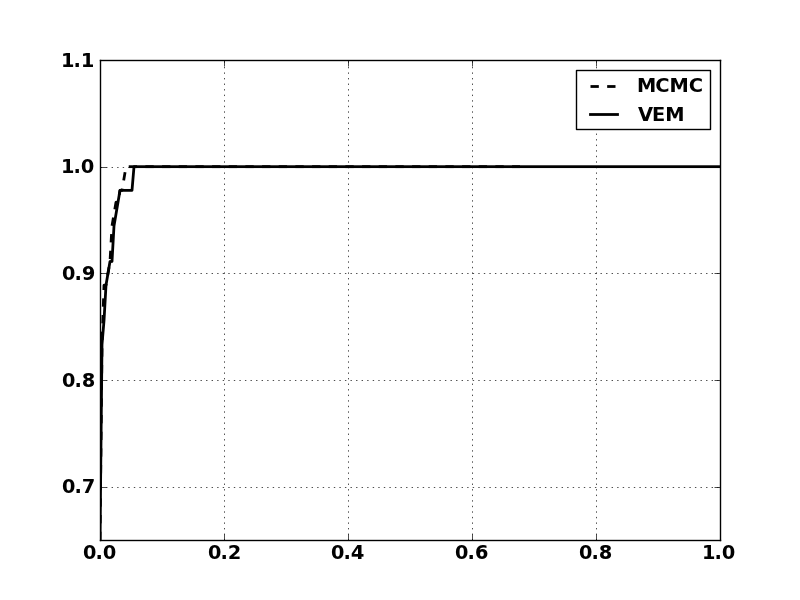}&
\rotatebox{90}{ \hspace{1.1cm} \small sensitivity}&
\includegraphics[height=4cm, width=6cm, trim= 1.6cm 1cm 1.5cm 1.3cm, clip=true]{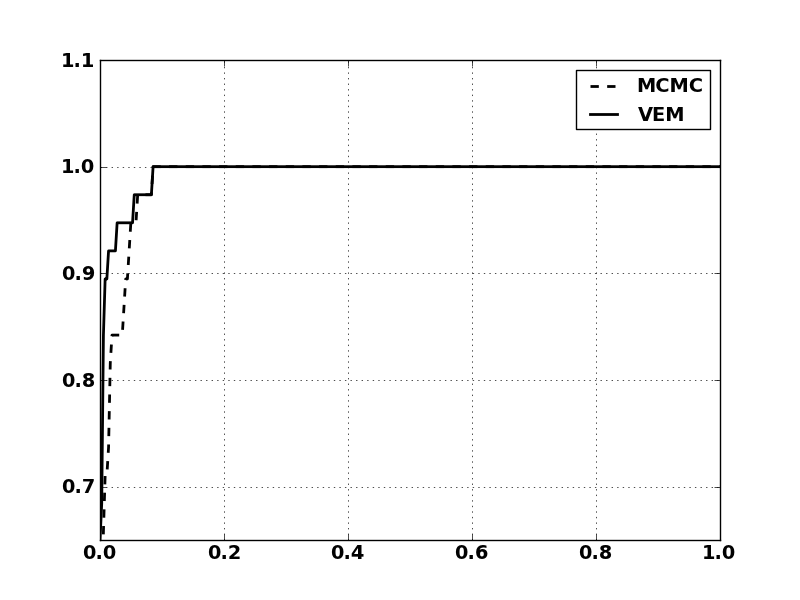} \\
&\small \raisebox{1.2cm}{1-specificity}&&\small \raisebox{1.2cm}{1-specificity}
\end{tabular}
\vspace{-1.5cm}
 \caption{ROC curves associated with the label estimates using VEM and MCMC.
Condition $m=1$ presents higher CNR than condition $m=2$. Curves are plotted in solid and dashed line for
VEM and MCMC, respectively.\label{fig:roc}}
\end{figure}

The posterior probability maps~(PPM) obtained using VEM and MCMC
are shown in Fig.~\ref{fig:labels}[middle] and
Fig.~\ref{fig:labels}[right]. PPMs here correspond to the activation class assignment probability.
These figures clearly show the gain
in robustness provided by the variational approximation. This
gain consists of lower miss-classification noise (a lower false
positive rate) illustrated by higher PPM values, especially for the experimental condition
with the lowest CNR ($m=2$). \\
For a  quantitative evaluation,  the ROC curves corresponding to
the estimated PPMs using both algorithms were computed. They are
reported in Fig.~\ref{fig:roc} and confirm that the proposed
VEM approach outperforms  the MCMC implementation for the second
experimental condition ($m=2$). Conversely, for the higher CNR
($m=1$), the MCMC approach performs slightly better.

\begin{figure}[!ht]
\centering
\hspace*{-.6cm}
\begin{tabular}{c c c c c |cc}
&  Ground Truth&  MCMC&  VEM&&VEM-MCMC&\\
\raisebox{1.5cm}{ $m=1$}&
\includegraphics[height=3cm,width=3cm]{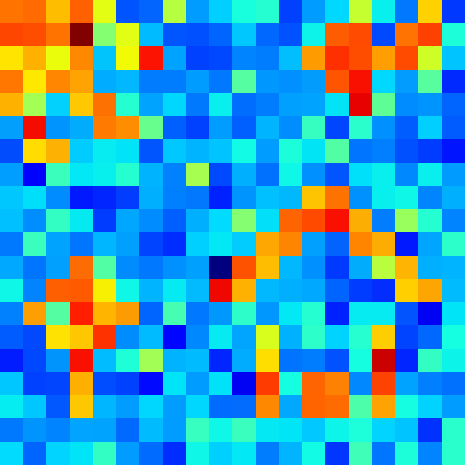}&
\includegraphics[height=3cm,width=3cm]{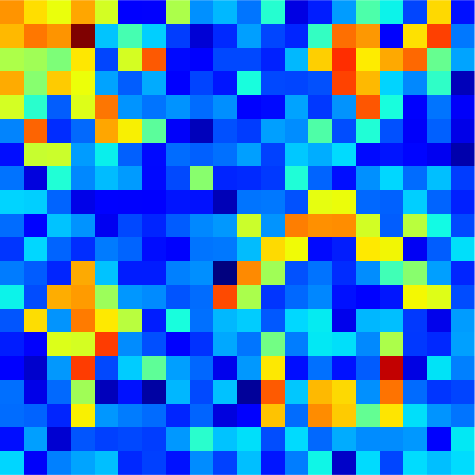}&
\includegraphics[height=3cm,width=3cm]{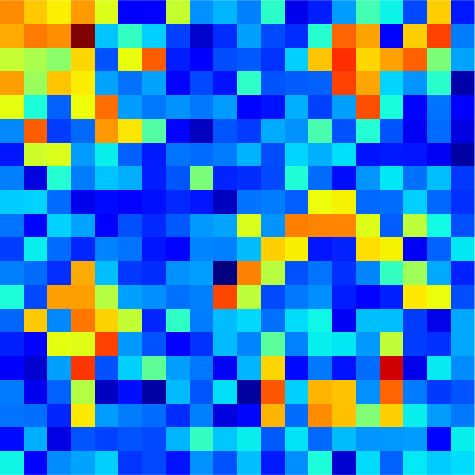}&
\includegraphics[height=3cm,width=.6cm]{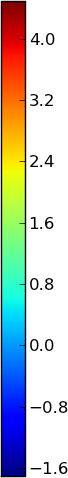}&
\includegraphics[height=3cm,width=3cm]{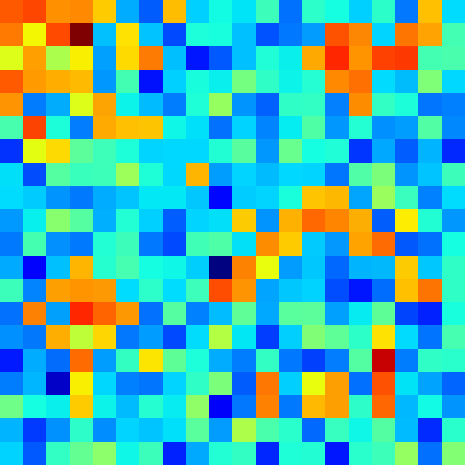}&
\includegraphics[height=3cm,width=.6cm]{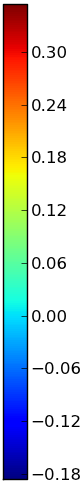}
\\
\raisebox{1.5cm}{ $m=2$}&%\hspace{-.8cm}
\includegraphics[height=3cm,width=3cm]{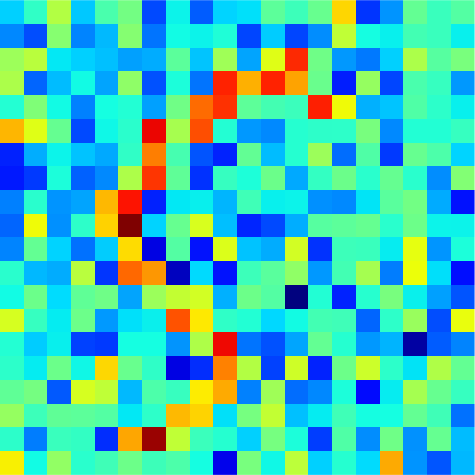}&
\includegraphics[height=3cm,width=3cm]{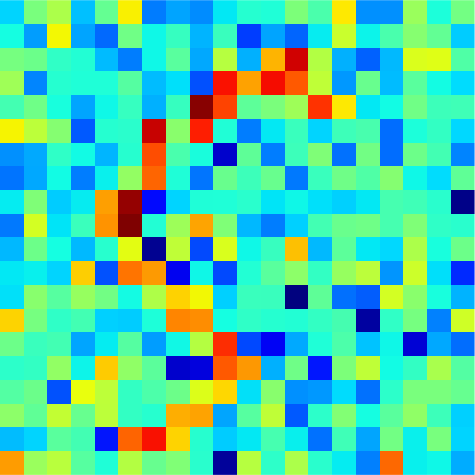}&
\includegraphics[height=3cm,width=3cm]{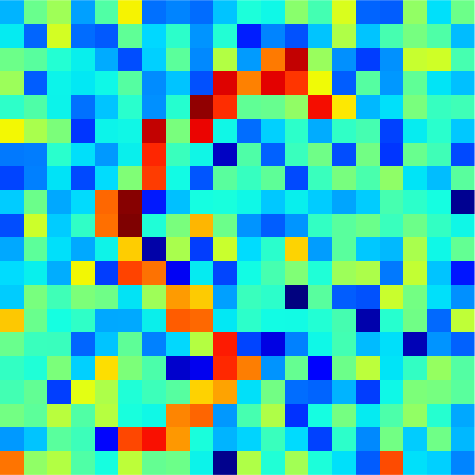}&
\includegraphics[height=3cm,width=.6cm]{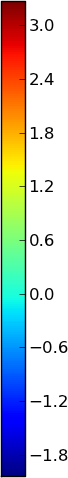}&
\includegraphics[height=3cm,width=3cm]{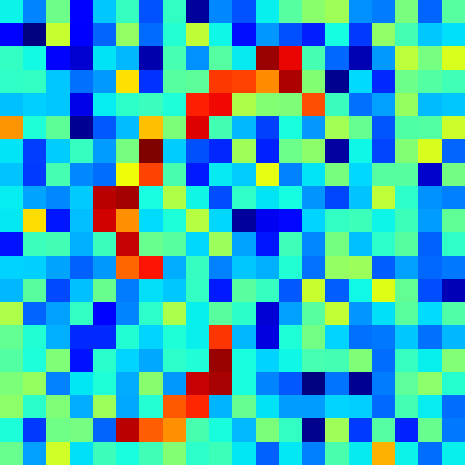}&
\includegraphics[height=3cm,width=.6cm]{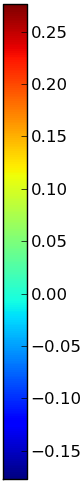}\\
\vspace{-.9cm}
\end{tabular}
\caption{ Ground truth and NRL estimates by MCMC and VEM~(first three images), and difference NRL image (right). \label{fig:nrls}}
\end{figure}

Fig.~\ref{fig:nrls} shows the NRL estimates obtained by the two methods. Although some differences
are exhibited on the PPM, both algorithms report similar qualitative results.
However, Fig.~\ref{fig:nrls}[right] shows the difference
between NRL estimates (VEM-MCMC). It is worth noticing in this figure that regions corresponding to activated areas for
the two conditions
present positive intensity values, which shows that VEM helps retrieving stronger NRL values for activated area compared to MCMC.
Quantitatively speaking, the gain in robustness is confirmed by reporting the Mean Square Error (MSE) values on NRL estimates
which are slightly lower
using VEM compared to MCMC for the first experimental condition ($m=1$: $\rm MSE_{\rm MCMC} = 0.012$
vs. $\rm MSE_{\rm VEM} = 0.010$), as well as for the second experimental condition ($m=2$: $\rm MSE_{\rm MCMC} = 0.010$
vs. $\rm MSE_{\rm VEM} = 0.009$). These error values indicate that, even though the MCMC algorithm gives the most precise PPMs for the
high CNR condition (Fig.~\ref{fig:roc}, $m=1$), the VEM approach is more
robust than its MCMC alternative in terms of estimated response levels. These values also indicate slightly lower MSE for the
second experimental conditions ($m=2$) compared to the first one ($m=1$) with higher CNR. This difference
is due to the presence of larger non-activated area
 used for $m=2$ where low NRL values are simulated, and for which MSE is very low.

\begin{figure}[!ht]
\centering
\rotatebox{90}{ \hspace{0.8cm} \small $\%\Delta $BOLD signal}
\includegraphics[height=4.5cm,width=7cm]{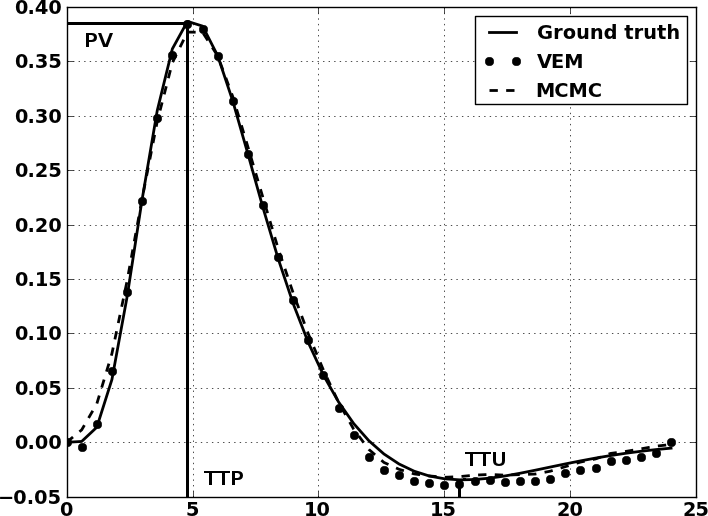}\\
\small Time~(s)
\vspace{-.6cm}
 \caption{Ground truth and estimates HRFs using the MCMC and VEM algorithms. \label{fig:hrfs1}}
\end{figure}

As regards HRF estimates, Fig.~\ref{fig:hrfs1} shows both retrieved shapes using MCMC and VEM. Compared to the ground truth curve
(solid line),  the two approaches give very similar results and
preserve the most important features of the original HRF like the
peak value~(PV), time-to-peak~(TTP) and time-to-undershoot~(TTU).

\subsubsection{Estimation robustness}\hfill \\
Since estimation errors may be caused by several sources of
perturbation ({\it e.g.} noise level, stimulus
density,...), several experiments have been conducted with different values of simulation parameters.

\begin{figure}[!ht]
\centering
\begin{tabular}{c c c c}
&$m=1$ & &$m=2$\\
\rotatebox{90}{\hspace{2cm} \small MSE}&
\includegraphics[height=5cm,width=7.3cm]{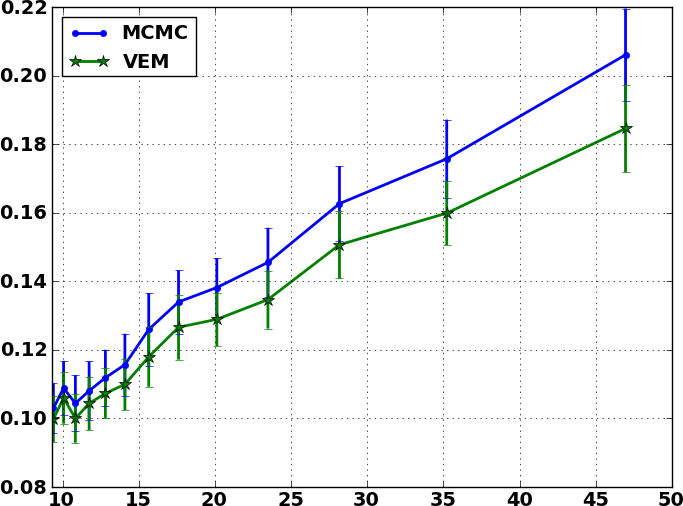}&
\rotatebox{90}{\hspace{2cm} \small MSE}&
\includegraphics[height=5cm,width=7.3cm]{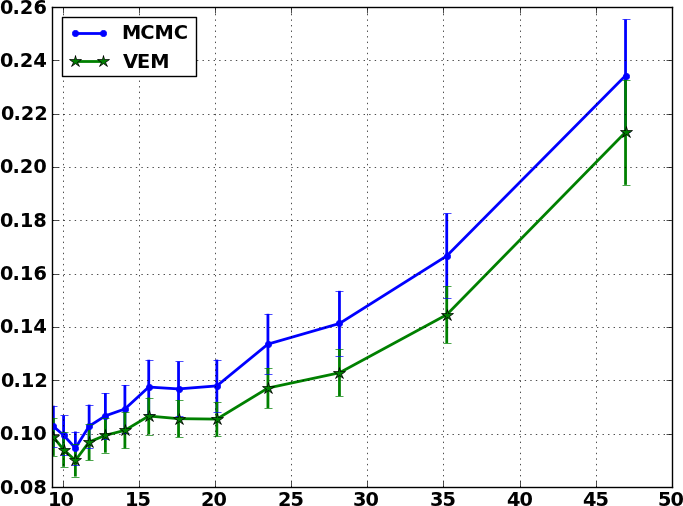}\\
&  \small \raisebox{1.2cm}{ISI~(s)}&&\small \raisebox{1.2cm}{ISI~(s)}
\end{tabular} \vspace{-1.4cm}
 \caption{MSE evolution of NRL estimates wrt Inter Stimuli Interval (ISI) for experimental conditions $m=1$ and
 $m=2$. Vertical bars represent empirical standard deviations
 computed over 100 simulations.
\label{fig:stimdens}}
\end{figure}

\paragraph{Varying the stimulus density}\hfill \\
In this experiment, several simulations have been generated using
different stimulus densities in the artificial paradigm~(from 5 to
30 stimuli), which leads to different Inter Stimuli Intervals (ISI) (from
47~s to 9~s). Note here that the stimuli are interleaved between the two conditions so that the above-mentioned
ISIs correspond to the time interval between two events irrespective to the condition they belong to.
A second order autoregressive noise~(AR(2)) has also
been used since it has been reported in the literature that such a
model provided realistic BOLD signal~\cite{Makni_06}. The rest of
the model is specified as before.
In order to quantitatively evaluate the robustness of the proposed VEM approach
to input Signal to Noise Ratio
(SNR)\footnote{The SNR is given by: ${\rm SNR} = 10 \log
\froc{\sum\limits_{j \in \Pc_\gamma } \|\Sb_j
\hb_\gamma\|^2}{\sum\limits_{j \in \Pc_\gamma
}\norm{\varepsilonb_j}^2}$.},
results (assuming white noise for both algorithms) are compared while varying the stimulation rate
during the BOLD signal acquisition.
% Analysis results (assuming white noise for both algorithms) are then
% quantitatively compared in order to evaluate the robustness of the
% proposed VEM approach to input Signal to Noise Ratio
% (SNR)\footnote{The SNR is given by: ${\rm SNR} = 10 \log
% \froc{\sum\limits_{j \in \Pc_\gamma } \|\Sb_j
% \hb_\gamma\|^2}{\sum\limits_{j \in \Pc_\gamma
% }\norm{\varepsilonb_j}^2}$.  } variations induced by varying the
% stimulation rate during the BOLD signal acquisition.
Fig.~\ref{fig:stimdens} illustrates the MSE evolution related to
the NRL estimates for both experimental conditions wrt the ISI
(or equivalently the stimulus density) in the experimental
para\-digm. This figure shows that at low SNR level, {\it i.e.}
high ISI (or low stimulus density),
 and for both conditions, VEM is more
robust to model mispecification. In contrast, at low ISI ({\it
i.e.} high stimulus density), the two me\-thods perform similarly
and remain quite robust. It should be noted here that, as reported
in Section~\ref{sec:perf}, the error
 values on NRL estimates remain comparable for both experimental conditions and for all ISI values, although PPM results may
present some imprecisions for the low CNR condition.
Results shown in Fig.~\ref{fig:stimdens} were obtained over
100 simulations to investigate the NRL estimation MSE
variance through runs. It can therefore be noticed through this
figure that, for the two experimental conditions, higher MSE variance (larger error bars)
is obtained with MCMC compared to VEM. Moreover, estimation variance across simulations (error bars) increase with ISI as expected.

\begin{figure}[!ht]
\centering
\begin{tabular}{c c c c}
&HRF&&PV\\
\rotatebox{90}{\hspace{2cm} \small MSE}&
\includegraphics[height=4.5cm,width=6.8cm]{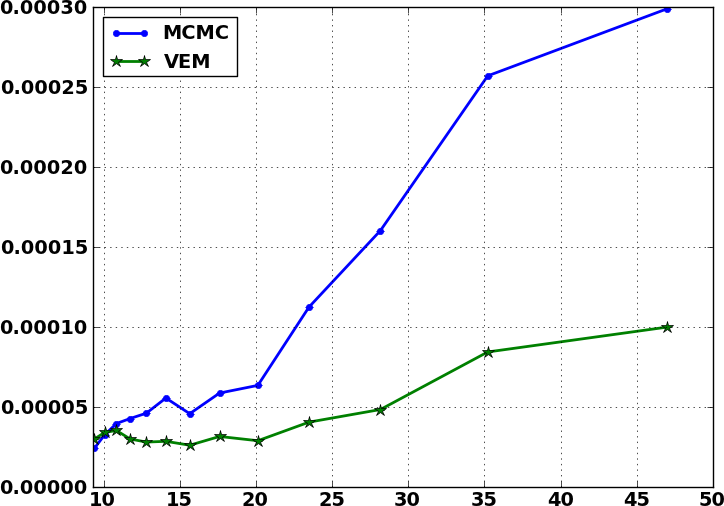}&
\rotatebox{90}{\hspace{2cm} \small MSE}&
\includegraphics[height=4.5cm,width=6.8cm]{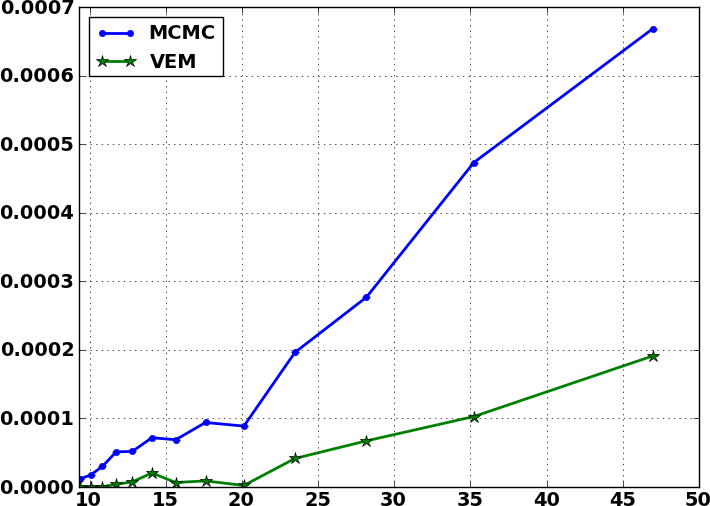}\\
&\small \raisebox{1.2cm}{ISI~(s)}&&\small \raisebox{1.2cm}{ISI~(s)}
\end{tabular}
\vspace{-1.4cm} \caption{MSE for the HRF estimates (left) and the
corresponding peak values (right) wrt interstimuli interval
expressed in seconds for VEM and MCMC.\label{fig:hrfs}}
\end{figure}

As regards hemodynamics properties, Fig.~\ref{fig:hrfs}[left] depicts
the MSE on HRF estimates inferred by the VEM and MCMC wrt the ISI
(or equivalently the stimulus density). The VEM approach performs
better than the MCMC one especially at low stimulus density (high
ISI values), but the error level remains relatively low for both
methods. When evaluating the estimation robustness of the key HRF
features~(PV, TTP and TTU), it turns out that the TTP and TTU estimates remain the same irrespective to the
inference algorithm, which
corroborates the robustness of the developed approach (results not shown).
As regards PV estimates, Fig.~\ref{fig:hrfs}[right] shows the MSE values
wrt the ISI. The VEM algorithm still performs better than the MCMC
one mainly at high ISI values (low stimulus density). For more
complete comparisons, similar experiments have been conducted
while changing the ground truth HRF properties~(PV, TTP, TTU), and
coherent results have been obtained.

\begin{figure}[!ht]
\centering
% \begin{tabular}{c c c c c}
\begin{tabular}{c c c c}
&$m=1$&&$m=2$\\%&\\
\rotatebox{90}{ \hspace{1.8cm} \small MSE}
&\includegraphics[height=4.5cm,width=6.5cm]{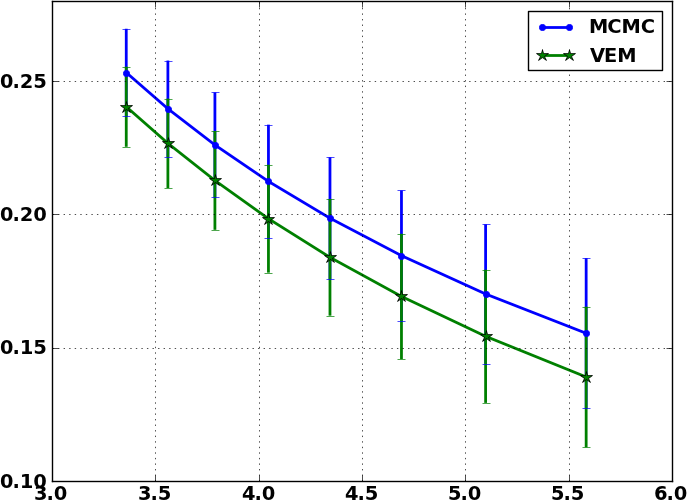}&
\rotatebox{90}{  \hspace{1.8cm} \small MSE}
&\includegraphics[height=4.5cm,width=6.5cm]{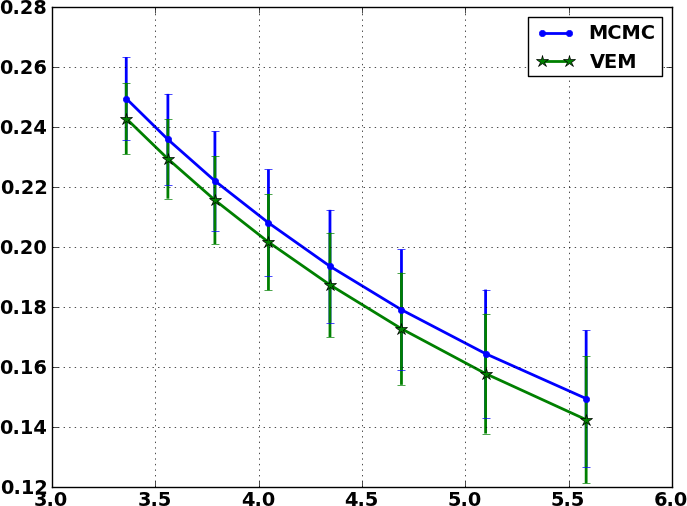}\\
&\small \raisebox{.8cm}{Input SNR~(dB)}&&\small \raisebox{.8cm}{Input SNR~(dB)}\\
\rotatebox{90}{ \hspace{1.8cm} \small MSE}&
\includegraphics[height=4.5cm,width=6.5cm]{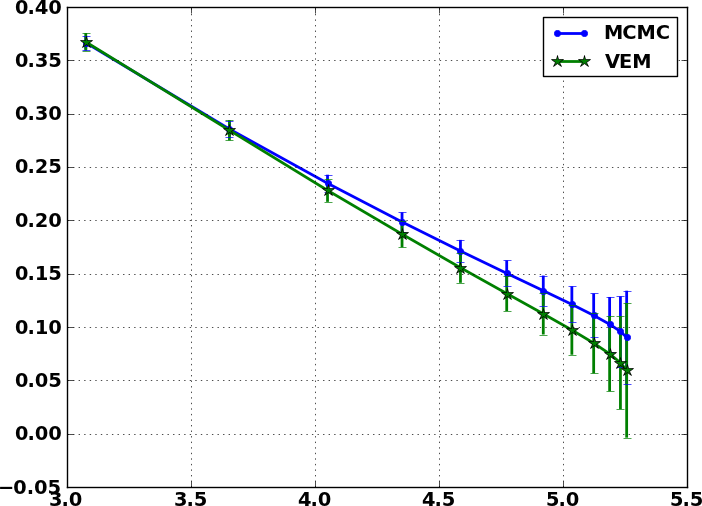}&\rotatebox{90}{ \hspace{1.8cm} \small MSE}
 &\includegraphics[height=4.5cm,width=6.5cm]{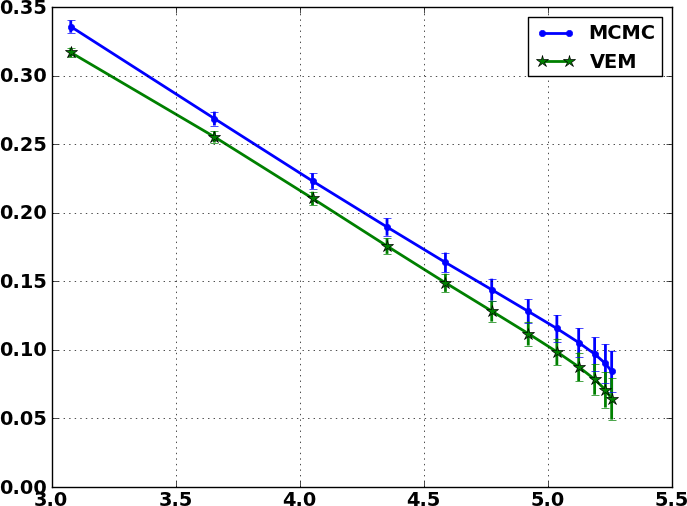}\\%& \raisebox{2cm}{(b)}\\
&\small \raisebox{.8cm}{Input SNR~(dB)}&&\small \raisebox{.8cm}{Input SNR~(dB)}%&
\end{tabular}
\vspace{-1.1cm}
 \caption{ MSE evolution of NRL estimates wrt input SNR~(AR(2) noise) by varying the noise variance~(top row) and the
amount of AR(2) noise autocorrelation~(bottom row). \label{fig:ARnoise}}
\end{figure}

\paragraph{Varying the noise parameters}\hfill \\
In this experiment, several simulations have been conducted using an AR(2) noise with different variance and correlation
parameters in order to illustrate the robustness of the proposed VEM approach to noise parameter fluctuation.
In Fig.~\ref{fig:ARnoise}, MSE on NRL estimates is
plotted against the input SNR when varying the noise variance
(Fig.~\ref{fig:ARnoise}[top]) and its amount of autocorrelation
(Fig.~\ref{fig:ARnoise}[bottom]). In the latter case, the AR parameters
are changed while maintaining a stable AR(2) process. As already
observed in~\cite{Casanova2008}, at a fixed input SNR value, the impact of
large autocorrelation is stronger than that of large noise
variance irrespective of the inference scheme. At low input SNR (as usually observed on real BOLD signals), this feature is
mainly shown through Figs.~\ref{fig:ARnoise}[bottom]-[top].
% As expected, these figures also show lower error values for the first experimental condition compared to the second one having lower input SNR.
Moreover, although a slight advantage is observed for the proposed VEM approach in
terms of MSE and for both experimental conditions, the two methods
 perform generally well with a relatively low error level.
The illustrated results were obtained over 100 simulations in
order to investigate the estimation error variance (vertical bars
in Fig.~\ref{fig:ARnoise}). These error bars show that changing the amount of correlation (Fig.~\ref{fig:ARnoise}[top])
induces lower variance across simulations than when changing
the noise variance (Fig.~\ref{fig:ARnoise}[bottom]).

% As already observed in~\cite{Casanova2008}, at low input
% SNR level, the impact of large autocorrelation is stronger than that of
% large noise variance irrespective of the inference scheme. This fact is emphasized by the lower MSE variance
% across simulations that is observed in Fig.~\ref{fig:ARnoise}[(b)] (varying the amount of autocorrelation) compared to the one observed in
% Fig.~\ref{fig:ARnoise}[(a)] (varying the noise variance). When comparing the two JDE versions, it turnes out that VEM performs slightly
% better by giving estimates of lower MSE.

%  For both experimental conditions, it
% appears that the MSE variance across runs is lower when varying
% the amount of autocorrelation than when varying the noise
% variance. This variance is also slightly lower with the proposed
% VEM approach compared to the MCMC one.

\begin{figure}[!ht]
\centering
\begin{tabular}{c cc c}
&$m=1$ & &$m=2$\\
\rotatebox{90}{ \hspace{1.2cm} \small estimated value}&
\includegraphics[height=4.5cm,width=6.3cm]{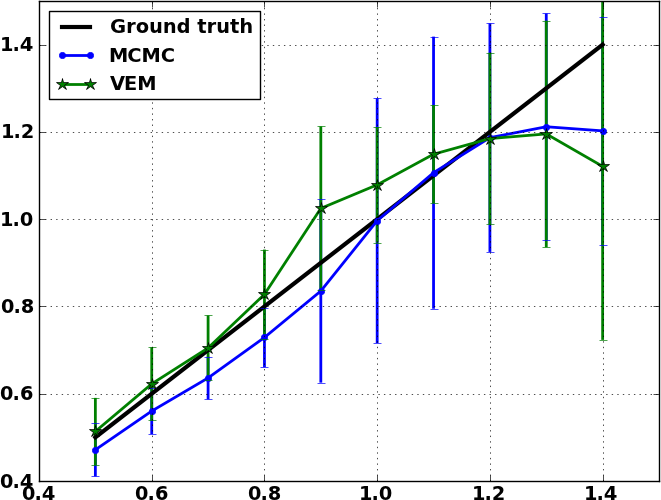}&
\rotatebox{90}{ \hspace{1.2cm} \small estimated value}&
\includegraphics[height=4.5cm,width=6.3cm]{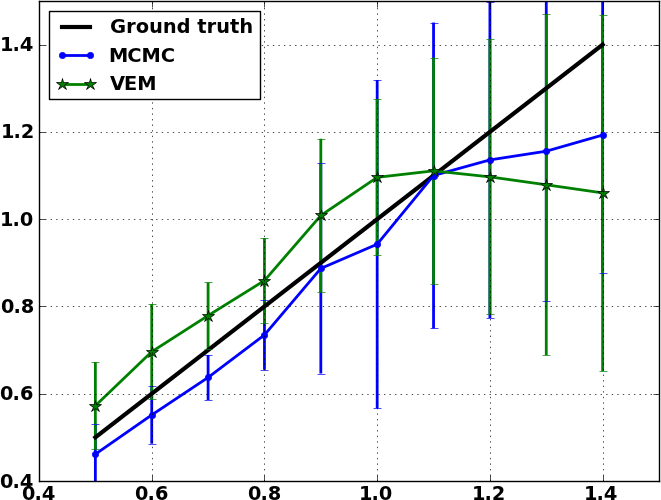}\\
&\small \raisebox{1.2cm}{real value}&&\small \raisebox{1.2cm}{real value}
\end{tabular}
\vspace{-1.5cm}
 \caption{Reference (black) and estimated mean values of $\betab$ VEM (green) MCMC (blue) for both experimental conditions
($m=1$ and $m=2$). Mean values and standard deviations (vertical
bars) are computed based on 100 simulations. \label{fig:betas}}
\end{figure}

\paragraph{Varying the spatial regularisation parameter}\hfill \\
This section is dedicated to studying the robustness of  the
spatial regularisation parameter estimation.
% In this section, the robustness of the estimatimation of the spatial regularisation parameter $\beta$ is evaluated.
When positive, this parameter favors  spatial regularity across
adjacent voxels, and hence  smoother activation maps. Fig.~\ref{fig:betas} shows the estimated mean
value and standard deviations for $\betab$ over 100 simulations
 using both algorithms and for the two experimental conditions. Three main regions can be distinguished for both experimental conditions.
The first one corresponds to $\beta$ valuer lower than $0.8$,
which approximatively matches the phase transition
critical value $\beta^c= \log(1 + \sqrt{2}) = 0.88$ for the 2-class Potts
model. For this region, Fig.~\ref{fig:betas} shows
that the VEM curve (green line) appears to be closer to the Ground
truth (black line) than the MCMC curve (blue line). Also, the
proposed VEM approach gives more precise estimation, especially
for the first experimental condition ($m=1$) having relatively
high CNR. The second region corresponds to $\beta$ values between
$0.8$ and $1.1$, where MCMC becomes more robust than VEM. Beyond
$\beta=1.1$, we can identify the third region where both methods
give less robust estimation than the first two regions. Based on
these regions, we conclude that the mean-field
variational approximation improves the estimation
performance up to a given critical value.
It turns out that such an approximation is more valid for low beta values, which usually correspond to $\beta$ values observed on real BOLD fMRI
data.\\
When comparing estimates for the two conditions, the curves in Fig.~\ref{fig:betas} show that both methods generally estimate more
precise $\beta$ values for the first experimental condition ($m=1$) having higher input CNR.
For both cases, and across the three regions identified hereabove, the  error bars show that the VEM approach generally gives less
scattered estimates (lower standard deviations) than the MCMC one, which confirms the
gain in robustness induced by the variational approximation.

% The VEM curve (green line) appears to be closer to the Ground
% truth (black line) than the MCMC curve (blue line). Also, the
% proposed VEM approach gives more precise estimation, especially
% for the first experimental condition ($m=1$) having relatively
% high CNR. However, Fig.~\ref{fig:betas}[right] also shows  that
% estimating $\betab$ becomes less precise for both methods when the
% input CNR decreases ($m=2$). For both cases, the  error bars
% clearly show that the VEM approach gives less scattered estimates
% (lower standard deviations) than the MCMC one, which confirms the
% gain in
% robustness due to the variational approximation.\\
% Then, for both algorithms and experimental conditions, the
% precision is clearly decreasing for $\beta$ values greater than
% the critical value $\beta^c= \log(1 + \sqrt{2}) = 0.88$
% corresponding to phase transition  for the Ising model (2
% classes).

% $\beta^c = \log(1 + \sqrt{2}) = 0.88$

Note here that estimated $\beta$ values in the experiment of Section~\ref{sec:perf} lie in the first region for the first experimental
condition ($\beta_1^{\rm MCMC}=0.74$ and $\beta_1^{\rm VEM}=0.75$). For the second condition ($m=2$),
and because input SNR is relatively low,
no clear conclusion can be made since MCMC and VEM give relatively different values
($\beta_2^{\rm MCMC}=0.61$ and $\beta_2^{\rm VEM}=1.01$) and no ground truth is available since the activation maps
have been drawn by hand and not simulated according to the Markov model.

\subsection{Real fMRI datasets}
\label{aubsec:real} 
\begin{figure}[!ht]
\center
\begin{tabular}{c@{~}c@{}c|}
% MCMC & VEM & \\
{\small MCMC} & {\small VEM} &\\
\includegraphics[height=3cm,width=4cm]{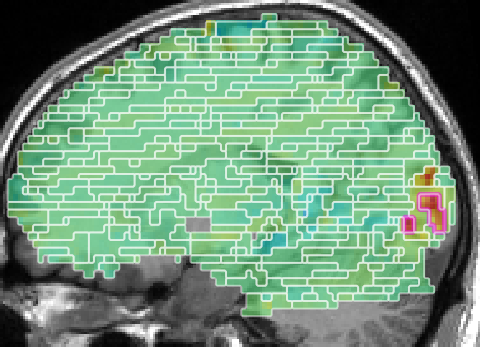} &
\includegraphics[height=3cm,width=4cm]{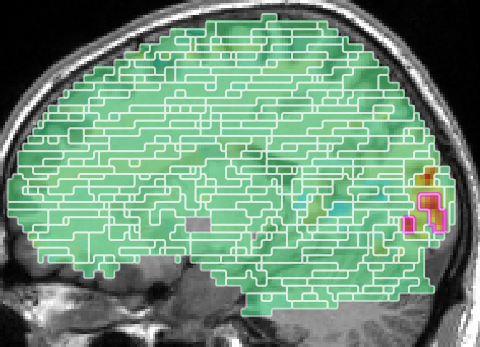} &
 \multirow{3}{*}{\includegraphics[height=6cm,width=1.cm]{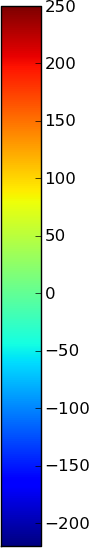}}
\\
\includegraphics[height=2.5cm,width=3cm]{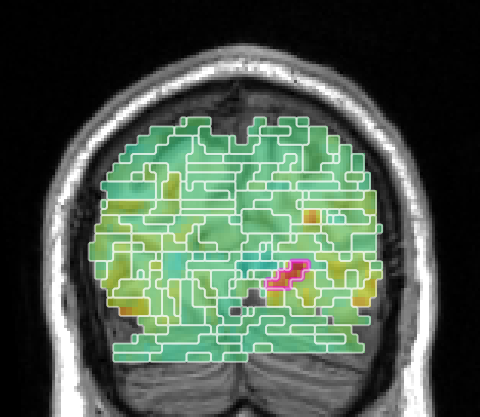} &
\includegraphics[height=2.5cm,width=3cm]{contrast_VA_MCMC_coronal_view.png}& \\
\includegraphics[height=3.4cm,width=3cm]{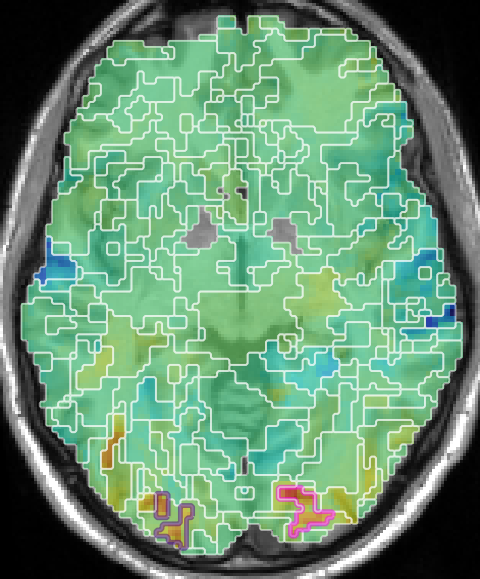} &
\includegraphics[height=3.4cm,width=3cm]{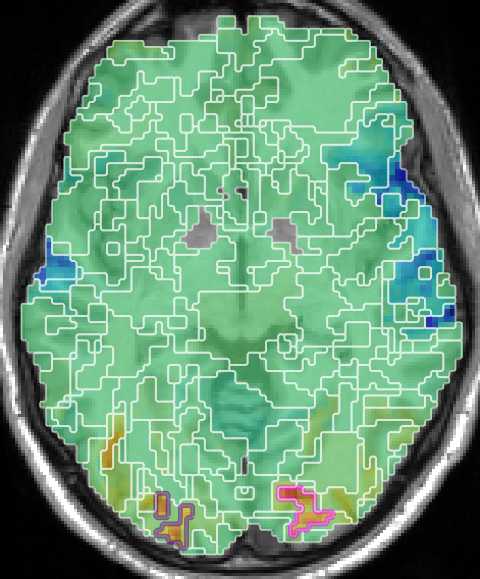}& \\
\end{tabular}
\begin{tabular}{c}
\textcolor{indigo}{(a)} \\
\includegraphics[height=4cm,width=6cm]{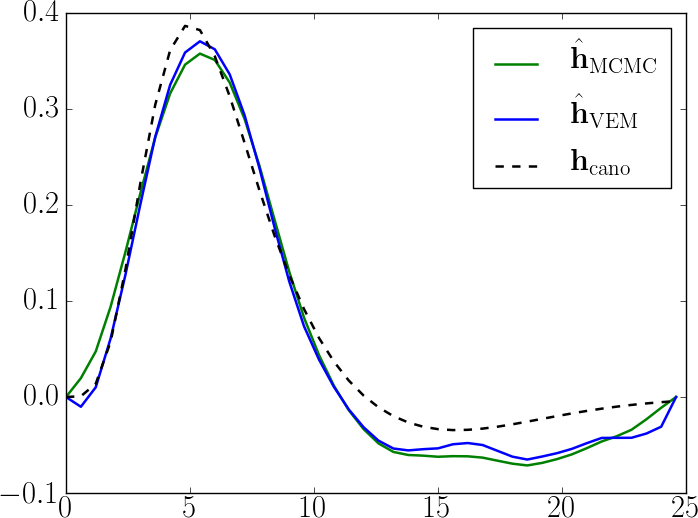} \\[-.3cm]
Time in sec. \\[.3cm]
\textcolor{magenta}{(b)} \\
\includegraphics[height=4cm,width=6cm]{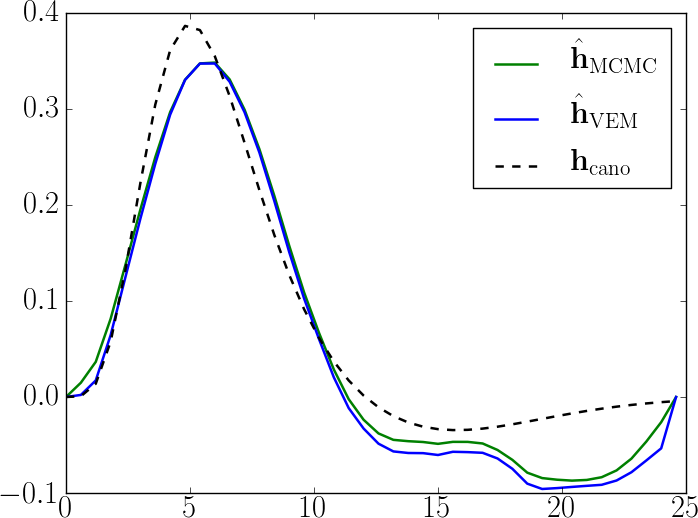} \\[-.3cm]
Time in sec. \\[.3cm]
\end{tabular}
\begin{tabular}{cc cc c}
\hline
$\beta^{\rm{vis.}}_{\rm MCMC}$ & $\beta^{\rm{aud.}}_{\rm MCMC}$ &
$\beta^{\rm{vis.}}_{\rm VEM}$ & $\beta^{\rm{aud.}}_{\rm VEM}$ \\
\includegraphics[height=3.4cm,width=3cm]{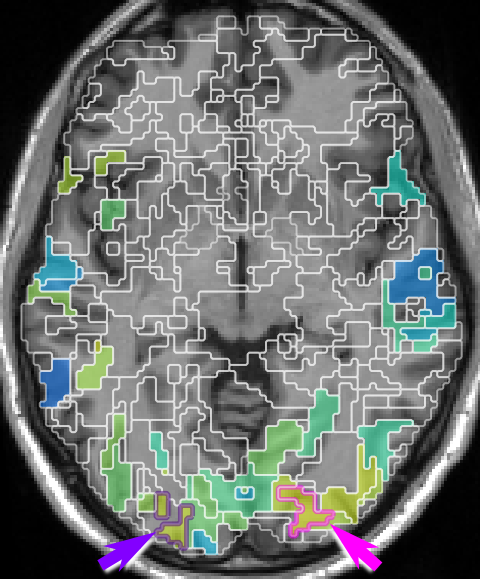}&
\includegraphics[height=3.4cm,width=3cm]{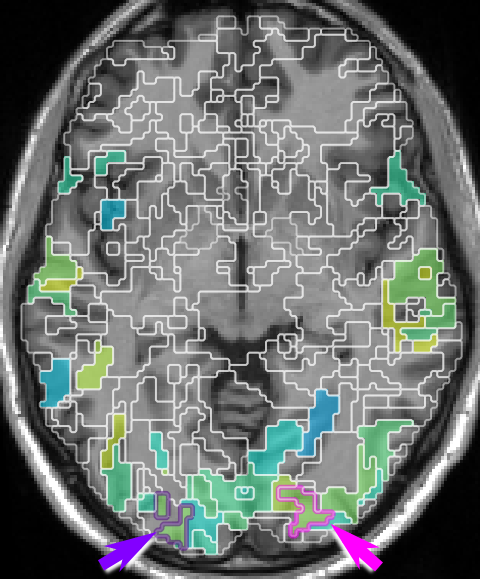}&
\includegraphics[height=3.4cm,width=3cm]{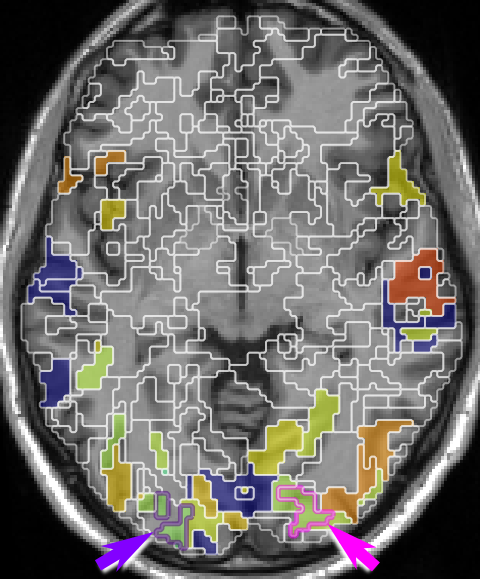}&
\includegraphics[height=3.4cm,width=3cm]{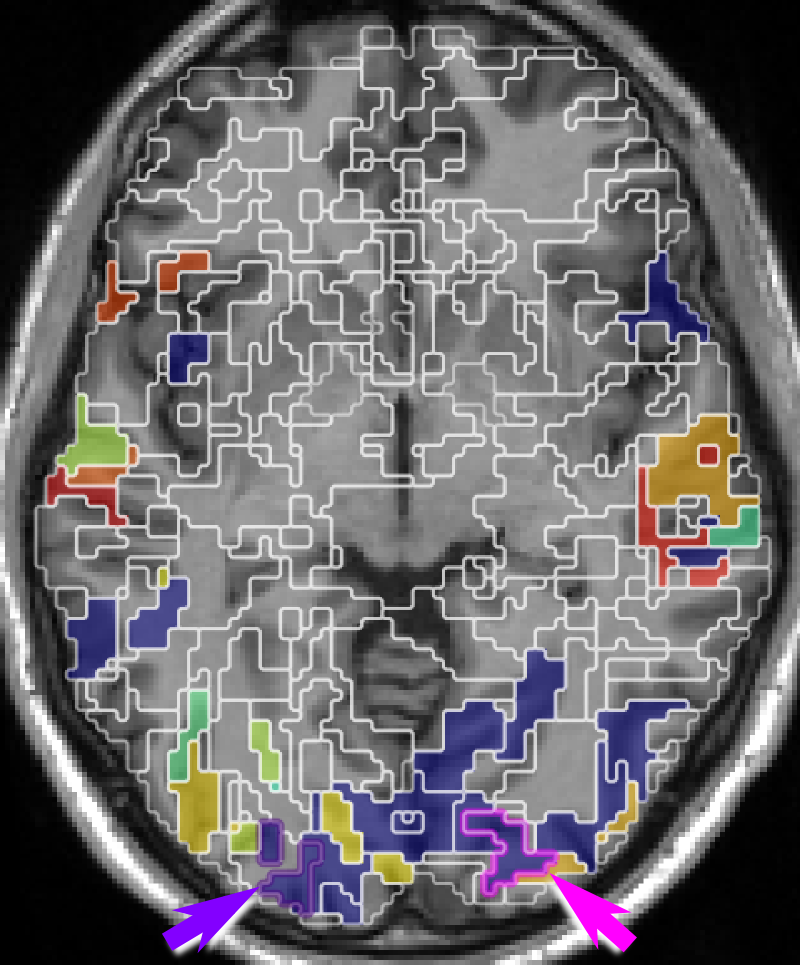}&
\includegraphics[height=3.5cm]{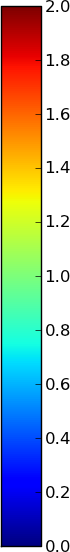}
\end{tabular}
\vspace{-.4cm} \caption{Results for the \textbf{Visual-Auditory}
contrast obtained by the VEM and MCMC JDE versions. Top left
column: contrast maps for MCMC, top middle column: contrast maps
for VEM, with sagittal, coronal and axial views from top to bottom
lines (neurological convention, left is left). On the top right
part: plots of HRF estimates for VEM and MCMC in the two regions
circled in \textcolor{indigo}{indigo} and
\textcolor{magenta}{magenta} on the maps: occipital
\textcolor{indigo}{left (a)} and \textcolor{magenta}{right (b)},
respectively. The canonical HRF shape is depicted in dashed line.
The bottom part shows axial maps of estimated regularisation
factors $\wh{\betab}$ for two conditions, auditory~(aud.) and
visual~(vis.), involved in the \textbf{VA} contrast. Parcels that
are not activated by any condition are hidden. For all maps, the
input parcellation is superimposed in white contours.
\label{fig:real_data_contrast_video_audio}}
\end{figure}

This section is dedicated to the experimental
validation of the proposed VEM approach in a real context.
Experiments were conducted on real fMRI datasets collected on
 healthy adult subjects who gave informed written consent. Data
were collected with a 3-Tesla Siemens Trio scanner using an MPRAGE
sequence for the anatomical MRI and a Gradient-Echo Echo Planar Imaging
(GRE-EPI) sequence
for the fMRI experiment. The acquisition parameters for the MPRAGE
sequence were set as follows: Time of Echo: $TE=2.98$ms; Time of Repetition: $TR=2300$ms; sagittal
orientation; spatial in-plane resolution: $1\times 1$mm$^2$;
Field of View: $\text{FOV}=256$mm$^2$ and slice thickness: $1.1$mm.
Regarding the EPI
sequence, we used the following settings: the fMRI session consisted
of $N = 128$ EPI volumes, where each scan was acquired using
$TR=2400$ms, $TE=30$ms, slice thickness: $3$mm, transversal
orientation, $\text{FOV}=192$mm$^2$ and spatial in-plane resolution was
set to $2\times 2$mm$^2$. Data was collected using a 32 channel
head coil to enable parallel imaging during the EPI acquisitions.
Parallel SENSE imaging was used to keep a reasonable Time of Repetition (TR) value in
the context of
high spatial resolution.\\
For the fMRI experiment, a functional localizer
paradigm~\cite{Pinel_07} was used, that enables to quickly map
cognitive brain functions such as reading, language comprehension
and mental calculations as well as primary sensory-motor
functions. It consists of a \emph{fast event-related} design
comprising sixty auditory, visual and motor stimuli, defined in
ten experimental conditions and divided in two presentation
modalities~(auditory and visual sentences, auditory and visual
calculations, left/right auditorily and visually induced motor
responses, horizontal and vertical checkerboards). The average ISI
is 3.75~s including all experimental conditions. After standard
pre-processing steps (slice-timing and motion corrections,
normalization to the MNI space), the whole brain fMRI data was
first parcellated into $\Gamma=600$ functionally homogeneous
parcels by resorting to the approach described
in~\cite{Thirion_06}. It consisted of a spatially constrained
hierarchical clustering (Euclidean distance, Ward's linkage) of
functional features extracted via a classical GLM analysis. This
parcellation was used as input of the JDE procedure, together with
the fMRI time series. We stress the fact that the latter signals
were not spatially smoothed prior to the analysis as opposed to
the classical SPM-based fMRI processing. In what follows, we
compare the MCMC and VEM versions of JDE by focusing on two
contrasts of interest: i) the \textbf{Visual-Auditory~(VA)}
contrast which targets positive and negative evoked activity in
the primary occipital and temporal cortices, respectively, and ii)
the \textbf{Computation-Sentences~(CS)} contrast which aims at
highlighting higher cognitive brain functions. Besides, results on
HRF estimates are reported for the two JDE versions and compared
to the canonical HRF, as well as maps of regularisation
factor estimates.\\
Fig.~\ref{fig:real_data_contrast_video_audio} shows results for the \textbf{VA} contrast. High positive values are bilaterally
 recovered in the occipital region and the overall cluster localizations are consistent for both MCMC and VEM algorithms. The only difference
 lies in the temporal auditory regions, especially on the right side, where VEM yields rather more negative values than MCMC.
VEM seems thus more sensitive than MCMC.
The bottom part of Fig.~\ref{fig:real_data_contrast_video_audio} compares the estimated values of the regularisation factors $\wh{\betab}$
between VEM and MCMC algorithms for two experimental conditions involved in the \textbf{VA}
contrast. Since these estimates are only relevant in parcels which are activated by at least one condition, a mask was applied to hide non-activated
parcels. We used the following criterion to classify a parcel as activated: $\max\{(\wh{\mu}_{1m})_{1 \leq m \leq M} \} \geq 8$ (and
non-activated otherwise). These maps of $\wh{\betab}$ estimates show that VEM yields more contrasted values between the visual and
auditory conditions. To be more precise, Table~\ref{table_beta_values_real_data}  provides the estimated $\wh{\betab}$ values in the
 highlighted parcels of interest. The auditory condition is not active and yields lower $\wh{\betab}$ values in both parcels
 whereas the visual condition is associated with higher values. The latter comment holds for both algorithms but VEM
 provides much lower values~($\wh{\betab}^{\rm aud.}_{\rm VEM}\approx 0.01$) than
 MCMC~($\wh{\betab}^{\rm aud.}_{\rm MCMC}\approx 1.07$) for the inactive condition.
For the active condition, the situation is comparable, with
$\wh{\betab}^{\rm vis.}_{\rm VEM} \approx 1.1$ and $\wh{\betab}^{\rm vis.}_{\rm MCMC} \approx 1.25$.
We illustrate here a noteworthy difference between VEM and
MCM and state the impact of the mean field and variational approximations, so that the hidden field has not
the same behaviour between the two algorithms. Still, this discrepancy is not visible on the NRL maps. \\
Fig.~\ref{fig:real_data_contrast_video_audio}[a-b] depicts HRF estimation results which are rather close for both methods
in the two regions under consideration. VEM and MCMC HRF estimations are also consistent with the canonical HRF shape.
Indeed, the latter has
been precisely calibrated on visual regions~\cite{Boynton96}.
We can note a higher variability in the undershoot part, which can
be explained first by the event-related nature of the paradigm
where successive evoked responses are likely to overlap so that it
is more difficult to disentangle their ends and second by the
signal strength which is inevitably low in the tail of the
response. To conclude on the \textbf{VA} contrast which focused on
well-known sensory regions, VEM provides sensitive results
consistent with the MCMC version, both wrt detection and estimation tasks.

\vspace{-.4cm}

\begin{figure}[!ht]
\begin{center}
\begin{tabular}{c@{~}c@{~}c|}
{\small MCMC} & {\small VEM} &\\
\includegraphics[height=3cm,width=4cm]{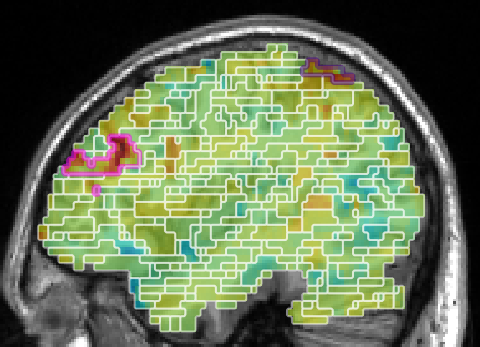} &
\includegraphics[height=3cm,width=4cm]{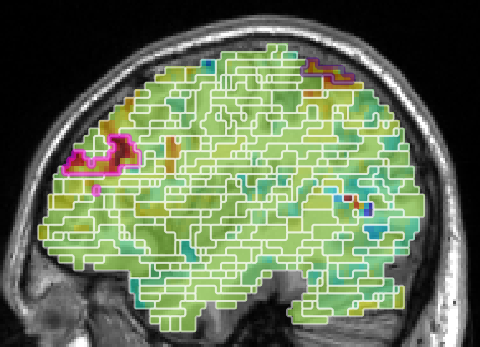} &
\multirow{3}{*}{\includegraphics[height=6cm,width=1cm]{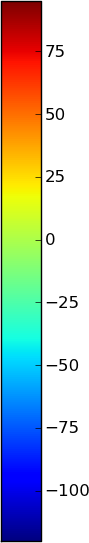}}\\
\includegraphics[height=2.5cm,width=3cm]{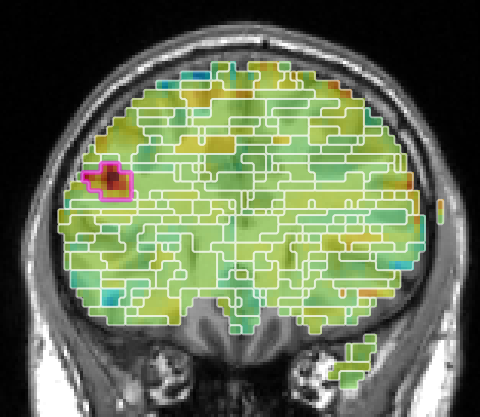} &
\includegraphics[height=2.5cm,width=3cm]{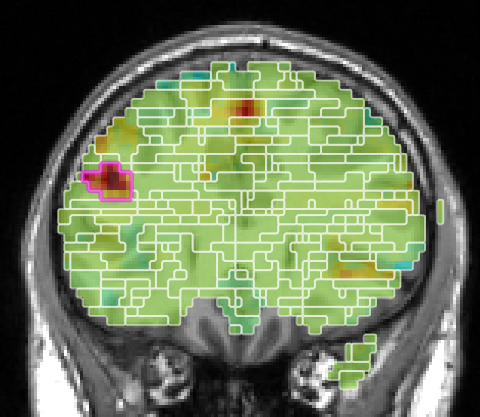} &\\
\includegraphics[height=3.4cm,width=3cm]{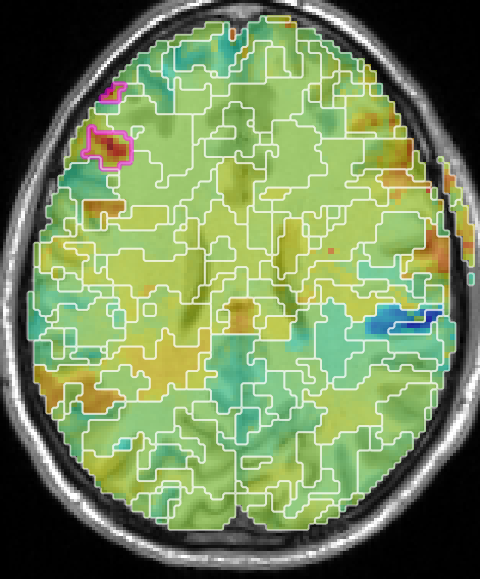} &
\includegraphics[height=3.4cm,width=3cm]{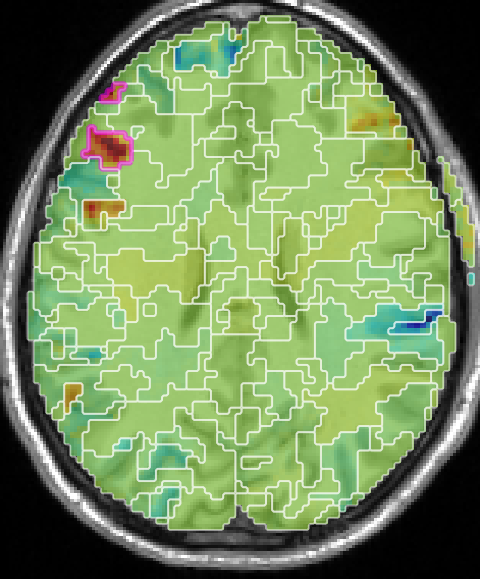} &\\
\end{tabular}
\begin{tabular}{c}
\textcolor{indigo}{(a)} \\
\includegraphics[width=5.4cm]{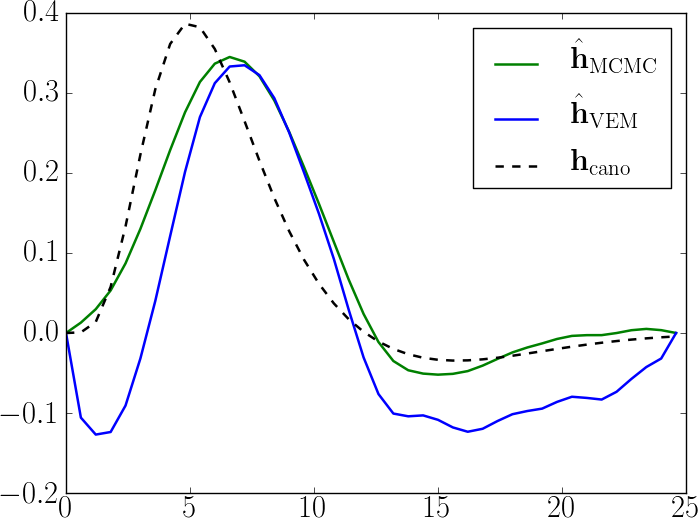}
 \\[-.3cm]
Time in sec.\\[.3cm]
\textcolor{magenta}{(b)} \\
\includegraphics[width=5.4cm]{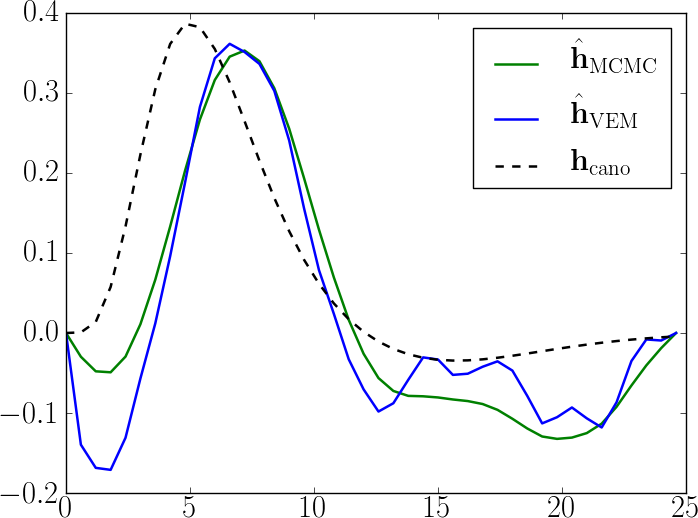}\\[-.3cm]
Time in sec.\\[.3cm]
\end{tabular}

\begin{tabular}{cc cc c}
\hline
$\beta^{\small{\rm comp.}}_{\rm MCMC}$ & $\beta^{\small{\rm sent.}}_{\small{\rm MCMC}}$ &
$\beta^{\small{\rm comp.}}_{\rm VEM}$ & $\beta^{\small{\rm sent.}}_{\small{\rm VEM}}$ \\
\includegraphics[height=3.4cm,width=3cm]{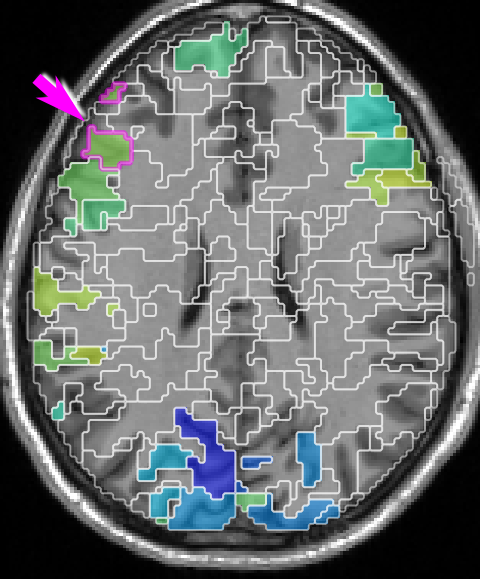}&
\includegraphics[height=3.4cm,width=3cm]{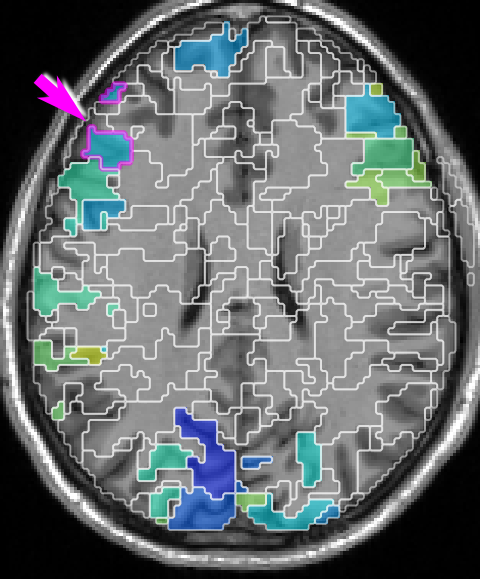}&
\includegraphics[height=3.4cm,width=3cm]{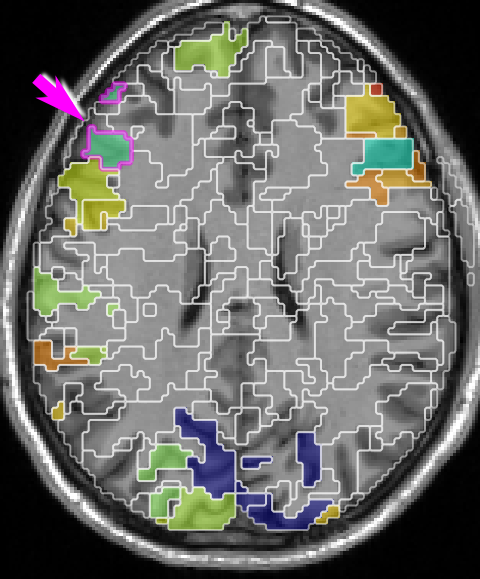}&
\includegraphics[height=3.4cm,width=3cm]{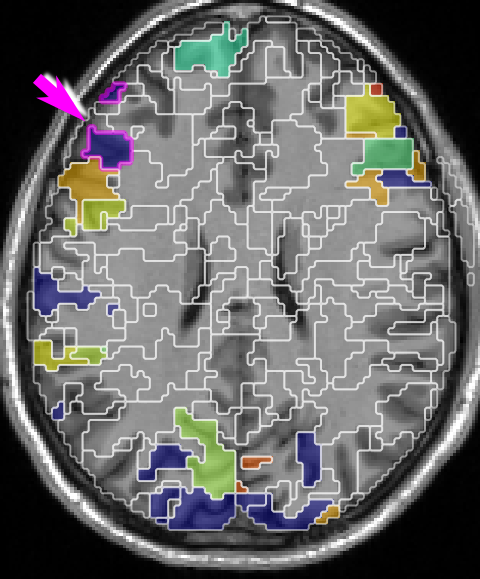}&
\includegraphics[height=3.5cm]{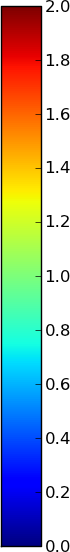}
\end{tabular}
\end{center}
\vspace{-.5cm} \caption{Results for the
\textbf{Computation-Sentences} contrast obtained by the VEM and
MCMC JDE versions. Top left column: contrast maps for MCMC, top
middle column: contrast maps for VEM, with sagittal, coronal and
axial views from top to bottom lines (neurological convention,
left is left). On the top right part: plots of HRF estimates for
VEM and MCMC in the two regions circled in
\textcolor{indigo}{indigo} and \textcolor{magenta}{magenta} on the
maps: \textcolor{indigo}{left parietal lobule (a)} and
\textcolor{magenta}{left middle frontal gyrus (b)}, repectively.
The canonical HRF shape is depicted in dashed line. The bottom
part shows axial maps of estimated regularisation factors
$\wh{\betab}$ for two conditions, computation~(comp.) and
sentence~(sent.), involved in the \textbf{CS} contrast. Parcels
that are not activated by any condition are hidden. For all
contrast maps, the input parcellation is superimposed in white
contours. \label{fig:real_data_contrast_CS}}
\end{figure}

Results related to the \textbf{Computation-Sentences (CS)}
contrast are depicted in Fig.~\ref{fig:real_data_contrast_CS}. As
for \textbf{VA}, contrast maps are roughly equivalent for VEM and
MCMC in terms of cluster localizations. Still, we observe that
MCMC seems quite less specific than VEM as positive values are
exhibited in the white matter for MCMC, and not for VEM (compare
especially the middle part of the axial views). For the estimates
of the regularisation factors, the situation is globally almost
the same as for the  \textbf{VA} contrast, with VEM yielding more
contrasted $\wh{\betab}$ maps than MCMC. However, these values are
slightly lower than the ones reported for the \textbf{VA}
contrast.

We first focus on the left frontal cluster, located in the middle
frontal gyrus which has consistently been exhibited as involved in
mental calculation~\cite{Gruber01}. HRF estimates in this region
are shown in Fig.~\ref{fig:real_data_contrast_CS}[(b)] and
strongly departs from the canonical version. Especially, the TTP
value is much more delayed with JDE (7.5 s), compared to the
canonical situation (5 s). The VEM and MCMC shapes are close to
each other, except for the beginning of the curves where VEM
presents an initial dip. This might be interpreted as a higher
temporal regularisation for the MCMC version. Still, the most
meaningful HRF features such as the TTP and the Full Width at Half
Maximum (FWHM) are very similar.

The second region of interest for the \textbf{CS} contrast is located in the inferior parietal lobule and is
also consistent with the computation task~\cite{Gruber01}. Note that the contrast value is lower than the one estimated
in the frontal region. Results for the regularisation factors, as shown in
Table~\ref{table_beta_values_real_data}~[$4^\text{th}$ col.], indicate that $\wh{\betab}$ for VEM and the
Sentence condition~(i.e, $\wh{\beta}_{\rm VEM}^{\rm sent.}$) is not as low as it was for the other parcels and the inactive conditions ($1.19$ against $0.01$).
This is due to the fact that both the Computation and the
Sentence conditions yield activations in this parcel, which is confirmed by the low contrast value.\\
HRF estimates are shown in Fig.~\ref{fig:real_data_contrast_CS}[(a)].
The statement relative to the previous region holds again: they strongly differ from the canonical version.
When comparing MCMC and VEM, even if the global shape and the TTP position are similar, the initial dip is still stronger
with VEM and the corresponding FWHM is also smaller than for the MCMC version.
As previously mentioned, this suggests that MCMC may tend to over-smooth the HRF shape.

The studied contrasts represent decreasing CNR situations, with the \textbf{VA} contrast being the stronger and
\textbf{CS} the weaker. From the detection point of view, the contrast maps are very similar for both JDE versions
and this result is only dimly affected by the CNR variation.
In contrast, HRF estimation results are much more sensitive to this CNR variation, with stronger discrepancies between the VEM and MCMC versions, especially for the HRF estimates associated with the \textbf{CS} parietal cluster. The latter
shows the weaker contrast amplitude. Still, both versions provide results in agreement on the time-to-peak and FHWM values. Indeed, the differences mainly concern the heading and tailing parts of the
HRF curves.
\vspace{-.4cm}

%% VA
%% beta phraseaudio roi 250 (magenta) MCMC =  1.07606
%% beta phraseaudio roi 468 (indigo) MCMC =  1.04671
%% beta phrasevideo roi 250 (magenta) MCMC =  1.28174
%% beta phrasevideo roi 468 (indigo) MCMC =  1.23896

%% beta phraseaudio roi 250 (magenta) VEM =  0.01
%% beta phraseaudio roi 468 (indigo) VEM =  0.01
%% beta phrasevideo roi 250 (magenta) VEM =  1.1408
%% beta phrasevideo roi 468 (indigo) VEM =  1.07635

%% CS
%% beta calculaudio roi 570 (magenta) MCMC =  1.08275
%% beta calculaudio roi 563 (indigo) MCMC =  1.07382
%% beta phraseaudio roi 570 (magenta) MCMC =  0.682583
%% beta phraseaudio roi 563 (indigo) MCMC =  0.640704

%% beta calculaudio roi 570 (magenta) VEM =  0.90655
%% beta calculaudio roi 563 (indigo) VEM =  0.822337
%% beta phraseaudio roi 570 (magenta) VEM =  0.01
%% beta phraseaudio roi 563 (indigo) VEM =  1.19151

\begin{table}
\caption{Comparison between JDE VEM and MCMC on the estimated regularisation parameters $\wh{\betab}$ for the experimental
conditions involved in the studied contrasts: \textbf{Visual-Auditive (VA)} and \textbf{Computation-Sentences (CS)}.
Results are provided for the two highlighted parcels for each contrast
(see Figs.~\ref{fig:real_data_contrast_video_audio} and \ref{fig:real_data_contrast_CS}).}
\centering
\label{table_beta_values_real_data}
\vspace{-.5cm}
\begin{tabular}{c cc|cc|cc|cc}
%\cline{2-5}
\multicolumn{1}{c}{}&\multicolumn{4}{c|}{\textbf{VA} contrast}&
\multicolumn{4}{c}{\textbf{CS} contrast}\\
\multicolumn{1}{c}{}&
\multicolumn{2}{c|}{\textcolor{magenta}{parcel $\gamma^\text{VA}_1$}} &
\multicolumn{2}{c|}{\textcolor{indigo}{parcel $\gamma^\text{VA}_2$}} &
\multicolumn{2}{c|}{\textcolor{magenta}{parcel $\gamma^\text{CS}_1$}} &
\multicolumn{2}{c}{\textcolor{indigo}{parcel $\gamma^\text{CS}_2$}} \\
\multicolumn{1}{c}{}& Vis. & Aud. & Vis. & Aud. & Comp. & Sent. & Comp. & Sent.\\
%\cline{1-1}
\hline
      %------------ VA --------------%  %---------------- CS ------------%
      %mag_vi %mag_au %ind_vi %ind_au   %mag_co  %mag_se %ind_co  %ind_se
MCMC & 1.28   & 1.08  & 1.24  & 1.05    & 1.08   & 0.68  & 1.07   & 0.64 \\
VEM  & 1.14   & 0.01  & 1.08  & 0.01    & 0.91   & 0.01  & 0.82   & 1.19 \\
\hline
\end{tabular}
\end{table}
% \rem{
% % \vspace*{-0.7cm}
% \begin{figure}[!ht]
% \centering
% \begin{tabular}{c c c}
% &{\tiny  Axial}&{\tiny  Sagittal}\\
% \raisebox{0.8cm}{\tiny MCMC} \hspace{.3cm}
% &\includegraphics[height=2cm,width=1.6cm]{con_calcul_phrase_mcmc_axial.png}&
% \includegraphics[height=2.cm]{con_calcul_phrase_mcmc_coronal.png}
% \includegraphics[height=2.cm]{color_bar_contrast.png} \\
% \raisebox{0.8cm}{\tiny
% VEM}\hspace{.3cm}&\includegraphics[height=2cm,width=1.6cm]{con_calcul_phrase_vem_axial.png}&
% \includegraphics[height=2.cm]{con_calcul_phrase_vem_coronal.png}\includegraphics[height=2.cm]{color_bar_contrast.png} \\
% \end{tabular}\hspace{1cm}
% % \begin{tabular}{c@{}c}
% \begin{tabular}{c}
% \raisebox{-2.75cm}{\includegraphics[height=2.3cm,width=4cm]{ehrf_MCMC_VEM_roi_285.png}}\hspace{-4.25cm}
% \raisebox{-3.25cm}{\rotatebox{90}{\hspace*{.6cm}\tiny{\%$\Delta$ BOLD signal}}}\hspace*{3cm} \\[-1.25cm]%&
% \raisebox{-2.75cm}{\includegraphics[height=2.3cm,width=4cm]{ehrf_MCMC_VEM_roi_274.png}}\hspace{-4.25cm}
% \raisebox{-3.25cm}{\rotatebox{90}{\hspace*{.6cm}\tiny{\%$\Delta$ BOLD signal}}}\hspace*{3cm} \\[-.5cm]%&
%  \hspace{-0.cm}{\tiny Time (s)}
% \end{tabular}%\vspace*{-.35cm}
% \caption{  {\bf Left:} Estimated contrast {\bf
% Computation-Sentences} by MCMC and VEM; {\bf Right:} HRF estimates
% by MCMC~(\textcolor{green}{green}) and
% VEM~(\textcolor{blue}{blue}) at maximum intensity peak~(top) and
% in a neighboring parcel~(bottom). Canonical HRF with dashed
% line.\label{fig:real}}
% \end{figure}
% }

\section{Algorithmic efficiency}
\label{sec_algorithmics_efficiency}
% \subsubsection{Convergence analysis}\hfill \\

In this section, the computational performance of the two approaches is
compared on both artificial and real fMRI datasets. Both algorithms were implemented in Python and fully optimized by resorting to the efficient array operations of the Numpy library~\footnote{\url{http://numpy.scipy.org}} as well as C-extensions for the computationally intensive parts (eg, NRL sampling in MCMC or the E-Z step for VEM). Moreover, our implementation handled distributed computing resources as the JDE analysis consists of parcel-wise independent processings which can thus be performed in parallel. This code is available in the PyHRF package~\footnote{\url{http://www.pyhrf.org}}. \\
For both the VEM and MCMC algorithms, the same stopping criterion is used. This
criterion consists of simultaneously evaluating the online
relative variation of each estimate. In other words, for instance
for the estimated $\wh{\hb}_\gamma$, one has to check whether
$c_H=\frac{||\wh{\hb}_\gamma^{(r+1)} -
\wh{\hb}_\gamma^{(r)}||_2^2}{||\wh{\hb}_\gamma^{(r)}||_2^2} \leq
10^{-5}$. By evaluating a similar criterion $c_A$ for the
NRLs estimates, the algorithm is finally stopped once $c_H \leq
10^{-5}$ and $c_A \leq 10^{-5}$. For the MCMC algorithm, this
criterion is only computed after the burn-in period, when the
samples are assumed to be drawn from the target distribution. The burn-in period has been fixed manually based on
several \textit{a posteriori} controls of simulated chains
relative to different runs (here 1000 iterations).  More
sophisticated convergence monitoring techniques \cite{gelman_92}
should be used to stop the MCMC algorithm, but we chose the same
criterion as for the VEM to carry out  a more direct comparison.

\begin{figure}[!ht]
\centering
\begin{tabular}{c c c c}
&HRF&&NRL\\
\rotatebox{90}{ \hspace{2cm} \small $c_H$}
&\includegraphics[height=4.2cm,width=7cm]{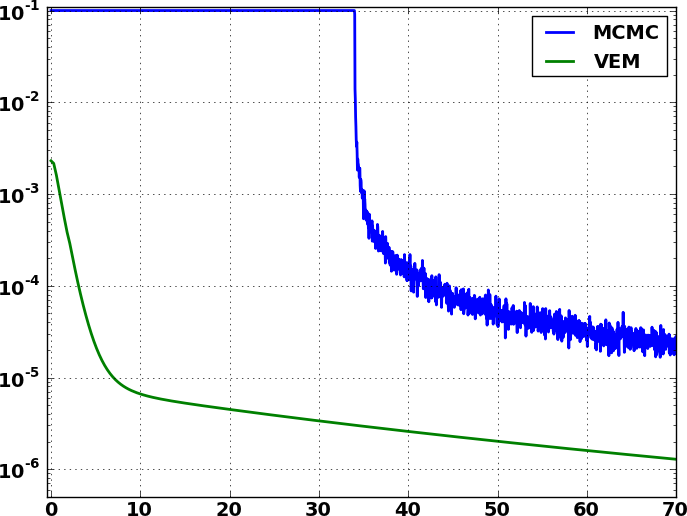}&
\rotatebox{90}{  \hspace{2cm} \small $c_A$}
&\includegraphics[height=4.2cm,width=7cm]{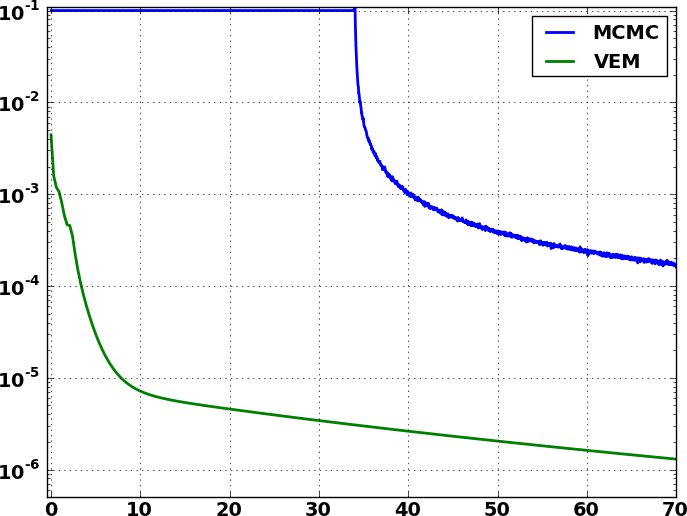}\\%&\raisebox{2cm}{VEM}\\
% &\tiny \raisebox{.8cm}{iteration number}&&\tiny \raisebox{.8cm}{iteration number}&\\
% \rotatebox{90}{ \hspace{1.5cm} \tiny $c_H$}&\includegraphics[height=4cm,width=6cm]{CV_mcmc_HRF.png}&
% \rotatebox{90}{ \hspace{1.5cm} \tiny $c_A$}
%  &\includegraphics[height=4cm,width=6cm]{CV_mcmc_NRL.png}& \raisebox{2cm}{MCMC}\\
&\small \raisebox{.8cm}{time (s)}&&\small \raisebox{.8cm}{time (s)}
\end{tabular}
\vspace{-1cm}
 \caption{ Convergence curves in logarithmic scale of HRF (left) and NRL (right) estimates using MCMC and VEM. 
% Plots here are given in a logarithmic scale. 
\label{fig:cv}}
\end{figure}

Considering  the artificial dataset presented  in
Section~\ref{sec:perf}, Fig.~\ref{fig:cv} illustrates the
evolution of $c_H$ and $c_A$ with respect to the computational
time for both algorithms. Only about 18 seconds are enough to
reach  convergence for the VEM algorithm, while the MCMC
alternative needs about 1 minute to converge on the same Intel
Core~4~-~3.20~GHz~-~4~Gb RAM architecture. The horizontal line in
the blue curve relative to the MCMC algorithm corresponds to the
burn-in period (1000 iterations).
% Convergence time for VEM and
% MCMC are indicated by green and blue arrows, respectively.
It can
also be observed through these curves that NRL estimates converge
faster than HRF estimates with the VEM approach, while the
convergence speed seems to be the same using the MCMC algorithm.
\vspace{-.4cm}
% Green and blue arrows indicate when the criteria $c_H$ and
% $c_A$ are fulfilled, respectively.

\begin{figure}[!ht]
\centering
% \hspace{-1.11cm}
\begin{tabular}{c c c c c c}
&(a)&&(b)&&(c)\\
\rotatebox{90}{\hspace{1.5cm} \small time~(s)}&
\hspace{-.3cm}\includegraphics[height=3.4cm,width=4.8cm]{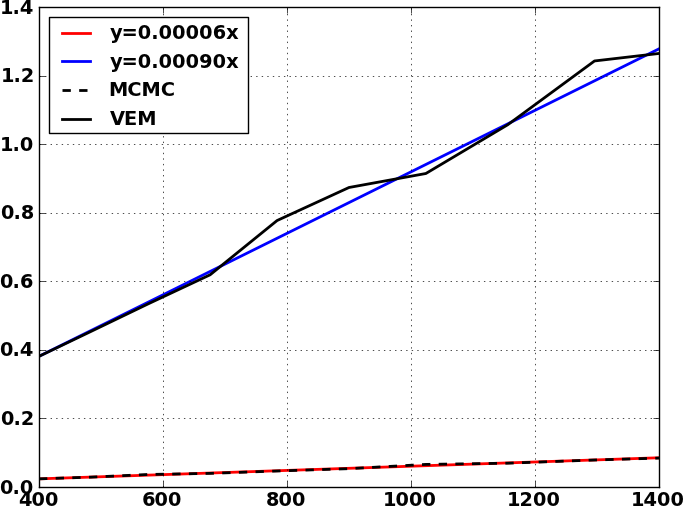}&
\rotatebox{90}{\hspace{1.5cm}\small time~(s)}&
\hspace{-.3cm}\includegraphics[height=3.4cm,width=4.8cm]{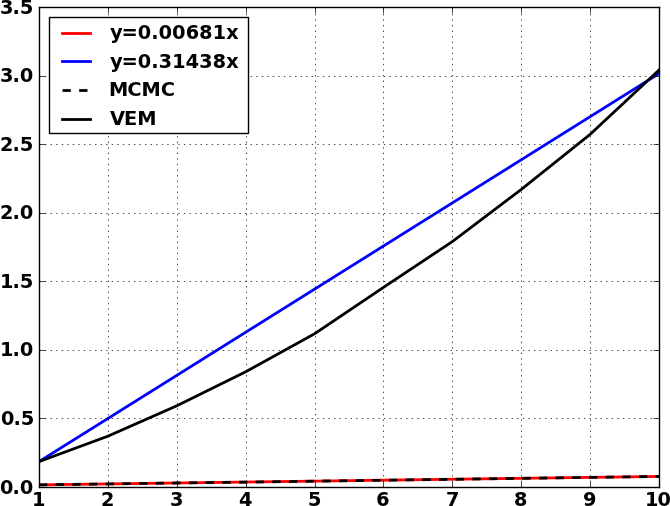}&
\rotatebox{90}{\hspace{1.5cm}\small time~(s)}&
\hspace{-.3cm}\includegraphics[height=3.4cm,width=4.8cm]{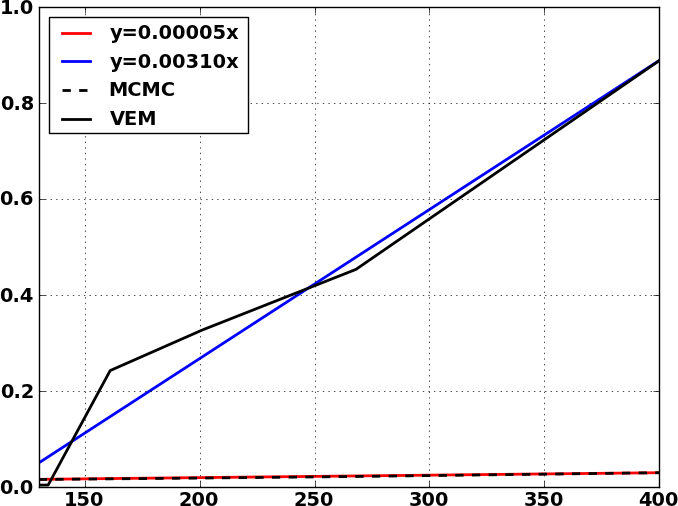}\\
&\small \raisebox{1.2cm}{number of voxels}&&\small \raisebox{1.2cm}{number of conditions}&&\small \raisebox{1.2cm}{number of scans}
\end{tabular}
\vspace{-1.8cm}
 \caption{ Evolution of the computational time per iteration using
the MCMC and VEM algorithms when varying the problem dimension according
to: (a): number of voxels; (b): number of conditions; (c): number of
scans.
 \label{fig:cvit}}
\end{figure}

\begin{figure}[!ht]
\centering
\btabu{ccc}
(a) & (b)& \\
\yaxis{$\Delta t$ in sec.}~\figc[height=5cm,width=6.7cm]{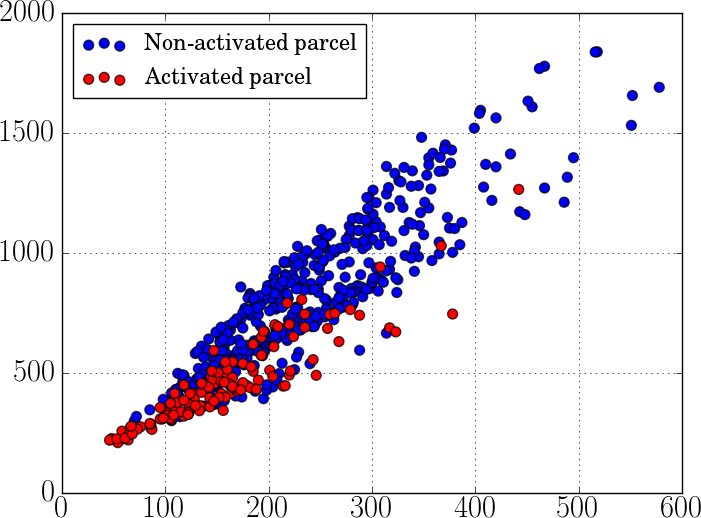}&
\yaxis{Gain factor}~\figc[height=4.8cm,width=6.7cm]{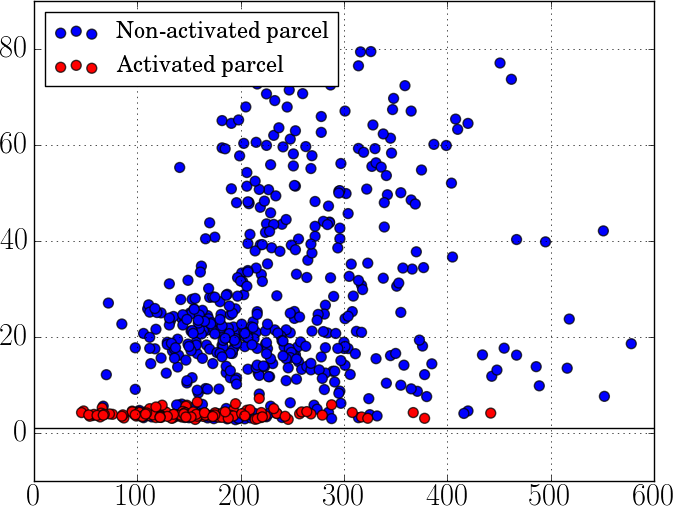}&\raisebox{-1.65cm}{\hspace{-.8cm}1}
\\[-.4cm]
{\small parcel size}&{\small parcel size} %\\%[.3cm]
\etabu
\vspace{-.4cm}
\caption{
Comparison of durations for MCMC and VEM analyses in terms of parcel size. \textbf{(a)}: differential timing
$t_\text{MCMC}-t_\text{VEM}$. \textbf{(b)}: gain factor of VEM compared to MCMC ($t_\text{MCMC}/t_\text{VEM}$), the horizontal line
indicates a gain factor of one ($t_\text{MCMC}=t_\text{VEM}$).
Circles are \textcolor{red}{red}-colored for parcels estimated as activated, ie
$\max\{(\wh{\mu}_{1m})_{1 \leq m \leq M} \} \geq 8$ and \textcolor{blue}{blue}-colored otherwise.
 \label{fig:profiling_parcels}}
\end{figure}

To illustrate the impact of the problem dimensions on the
computational cost of both methods, Fig.~\ref{fig:cvit} shows the
evolution of the computational time of one iteration when varying
the number of voxel (left), the number of experimental conditions
(middle) and the number of scans (right). The three curves
show that the computational time increases  almost linearly (see
the blue and red curves) for both algorithms, but with different
slopes. Blue curves (VEM) have steeper slopes than red ones (MCMC)
in the three plots showing that the computational time of one
iteration increases faster with VEM than with MCMC wrt the problem
dimensions.
\vspace{-.4cm}
\vspace{-.4cm}
\noindent As regards computational performance on the real fMRI data set presented in Section~\ref{aubsec:real} and comprising 600 parcels, the VEM also appeared faster as it took 1 hour 30 to perform a whole brain analysis whereas the MCMC version took 12 hours. These analysis timings were obtained by a serial processing of all parcels for both approaches. When resorting to the distributed implementation, the analysis durations boiled down to 7 mins for VEM and 20 mins for MCMC (on a 128-cores cluster).
To go further, we illustrate the computational
time difference ($t_\text{MCMC}-t_\text{VEM}$) between both
algorithms in terms of parcel size which ranged from 50 to 580 voxels.
As VEM vs. MCMC efficiency appears to be influenced by the level
of activity within the parcel, we resorted to the same criterion as in Section~\ref{aubsec:real} to distinguish non-active from active parcels and tag the analysis durations accordingly in Fig.~\ref{fig:profiling_parcels}.

Fig.~\ref{fig:profiling_parcels}[(a)] clearly shows that the
differential timing between both algorithms is higher for
non-activated parcels (blue dots) and increases with the parcel size, which
confirms the utility of the proposed VEM approach especially in
low CNR/SNR circumstances. To further investigate the gain in
terms of computational time induced by using the VEM approach,
Fig.~\ref{fig:profiling_parcels}[(b)] illustrates the gain factor
($t_\text{MCMC}/t_\text{VEM}$)  for activated and non-activated
parcels. This figure shows that the VEM algorithm always performs
better than the MCMC one since all obtained gain factors are
greater than 1 (see horizontal line in Fig.~\ref{fig:profiling_parcels}[(b)]). Moreover,  the gain factor is clearly higher for
non-activated parcels for which the input SNRs and CNRs are
relatively low, and it generally varies between 2.7 and 80.
\vspace{-.4cm}
\section{Conclusion and future work}\label{sec:conclusion}

In this paper, we have proposed a new method for parcel-based
joint detection-estimation of brain activity from fMRI data. The
proposed method relies on a Variational EM algorithm as an
alternative solution to intensive stochastic sampling used in
previous work \cite{Makni08}.
Compared to JDE MCMC, the proposed VEM approach does not require priors on the model parameters for inference
to be carried out. However, for more robustness and to
make the proposed approach completely auto-calibrated, the adopted
model may be extended by injecting additional priors on some of its parameters as detailed in Appendix~\ref{append:M} for
$\betab$ and $v_\hb$ estimation.\\
Illustrations on simulated and real
datasets have been deeply conducted in order to evaluate the
robustness of the proposed method compared to its MCMC counterpart
in different experimental contexts. Simulations have shown that
the proposed VEM algorithm gave more precise estimation of
activation labels and NRLs especially at low input CNR, while
giving similar
 performance for HRF estimation. Simulations have also shown that our approach
was more robust to stimulus density decrease (or equivalently ISI
time increase). Similar conclusions have been drawn wrt noise
level and autocorrelation structure. In addition, our VEM approach gave more
robust estimation of the spatial regularisation parameter and more
compact activation maps that are likely to better account for
functional homogeneity.  These good features of the VEM approach
are provided in shorter computational time than with the MCMC
implementation. Simulations have also been conducted to study the
computational time variation wrt the problem dimensions, which may
significantly vary from one experimental context to another. \\
Regarding real data experiments, VEM and MCMC showed similar
results
 with a higher specificity for the former. Although no ground
truth is available in this context, these results further
emphasize the interest of using VEM for a  high gain in
computational time. From a practical viewpoint, another advantage
of the proposed algorithm lies in its simplicity. By contrast to
the MCMC implementation of~\cite{Vincent10}, the VEM algorithm
only requires a simple stopping criterion. It is also more
flexible to account for more complex situations such as those
involving higher AR noise order, habituation modeling or
considering three instead of two activation classes with an
additional deactivation class.\\
To confirm the impact of the proposed approach, comparisons between
the MCMC and VEM approaches should also take place at the group
level. In other words, we should compare the results of random
effect analyses (RFX) based on a Student t-test on mean effects.
A first RFX analysis would correspond to the classical approach, in
which the input data are given by the normalized effects of a
standard individual SPM analysis. Subsequent analyses
would take the results of the VEM and MCMC approaches for each
subject as inputs to RFX analyses. In the same vein, a seminal study
has been performed in~\cite{Badillo11} where group results based on JDE MCMC intra-subject analyses provide
higher sensitivity than results based on GLM based intra-subject analyses.
%Obviously, such a study requires a multi-subjects parcellation~\cite{Thirion_06}.
Such a group-level validation would also shed
the light on the impact of the used variational approximation in
VEM. In fact, no preliminary spatial smoothing is used in the JDE
approach by contrast to standard fMRI analyses where this
smoothing helps retrieving clearer activation clusters. In this
context, the used mean field approximation especially at the E-Q
step should help getting less noisy activation clusters compared
to the MCMC approach.
% Other future work will focus on studying the impact of the fMRI
% data reconstruction algorithm \cite{Chaari_MEDIA_2011} on the JDE
% performance and investigate the impact of the variational
% approximation on the estimation results based on more or less
% noisy real fMRI data. \\
Eventually, akin to \cite{Makni08}, the
model used in our approach accounts for functional homogeneity at
the parcel scale. These parcels are assumed to be an input of the
proposed JDE procedure and can be \textit{a priori} provided
independently by any parcellation technique
\cite{Thirion_06,Tucholka08b}.
In the present work, parcels have been extracted based on functional features extracted via a classical GLM processing supposing a canonical HRF for
the entire brain. This assumption does not bias our HRF local model estimation since a large number of parcels is considered
($600$ parcels) with an average parcel size of $250$ voxels.\\
On real dataset, results may
therefore depend on the reliability of the used parcellation
technique. A sensitivity analysis has been performed in~\cite{Vincent08} on real data and for MCMC JDE version, that assesses the reliability of the used parcellation against a heavy approach where the parcellation was marginalised.
Still, it would be of interest to investigate the effect
of the  parcellation choice in the VEM context, and more generally to extend the
present framework to incorporate an automatic online parcellation
strategy to better fit the fMRI data while accounting for the HRF
variability across subjects, populations and experimental
contexts. The current variational framework has the advantage to
be easily augmented with  parcel estimation as an additional layer
in the hierarchical model.
This will then  raise the question of model selection, in
particular the issue of well separating parcels at
best {\it i.e.} in a sparse manner so as to capture the spatial
variability in hemodynamic territories while enabling the
reproducibility of parcel identification across fMRI datasets.
More generally, an approach to model selection can be easily
carried out within the VEM implementation as variational
approximations of standard information criteria based on penalised
log-evidence can be efficiently used~\cite{Forbes03}.

% \appendices
% \section{E-H step:}\label{append:E-H}

\appendix
\footnotesize
\subsection{E-H step:}\label{append:E-H}

For the E-H step,
 the expressions for
$\mb_{H_\gamma}^{(r)}$ and $\Sigmab_{H_\gamma}^{(r)}$ are:
\begin{align}
\Sigmab_{H_\gamma}^{(r)} &= [1/v_\hb^{(r-1)} \Rb\M + \sum_{j \in
\Pc_\gamma} \Bigpth{\sum_{m, m'} v_{A_{j}^mA^{m'}_{j}}^{(r-1)}
\Xb_{m}\T \Gammab_j^{(r-1)}
\Xb_{m'} + \wt{\Sb}_j\T\Gammab_j^{(r-1)} \wt{\Sb}_j}]\M\\
\mb_{H_\gamma}^{(r)} &= \Sigmab_{H_\gamma}^{(r)} \sum_{j \in
\Pc_\gamma}\wt{\Sb}_j\T \Gammab_j^{(r-1)} (\yb_j-
\Pb\ellb_j^{(r-1)}),
\end{align}
with  $\wt{\Sb}_j =\sum\limits_{m=1}^M m_{A_{j}^m }^{(r-1)}\Xb_m$, $m_{A_{j}^m }^{(r-1)}$ and $v_{A_{j}^m A_{j}^{m'}}^{(r-1)}$
denoting respectively the $m$  and  $(m,m')$  entries of the mean
vector ($\mb_{A_j}^{(r-1)}$) and covariance matrix
($\Sigmab_{A_j}^{(r-1)}$) of the current $\wt{p}_{A_j}^{(r-1)}$.

\subsection{E-A step:}\label{append:E-A}

The E-A step also leads to a Gaussian pdf  for
$\wt{p}^{(r)}_A$:
 $\wt{p}_A^{(r)} \sim \prod\limits_{j \in \Pc_\gamma} \Nc(\mb_{A_j}^{(r)},
\Sigmab_{A_j}^{(r)})$. The parameters are updated as follows:
\begin{align}\label{eq:varAvar}
\Sigmab_{A_j}^{(r)} = \Bigpth{\sum\limits_{i=1}^I \Deltab_{ij} +
\wt{\Hb}_{j}}\M &,&
\mb_{A_j}^{(r)} = \Sigmab_{A_j}^{(r)} \; \Bigpth{\sum\limits_{i=1}^I
\Deltab_{ij} \mub^{(r-1)}_i + \wt{\Gb}\T \Gammab_j^{(r-1)} (\yb_j-
\Pb \ellb_j^{(r-1)})} %\label{eq:meanAvar},
%\: \text{et}\:
\end{align}
\noindent  where a number of intermediate quantities need to be
specified. First, $\mub^{(r-1)}_i=\cro{\mu_{i1}^{(r-1)}, \ldots,
\mu_{iM}^{(r-1)}}\T$ and $\wt{\Gb}=
\Espud{\wt{p}_{H_\gamma}^{(r)}}{\Gb}$ where $\Gb$ is the matrix
$\Gb=\cro{\gb_1\I\ldots\I \gb_{M}}$ made of columns $\gb_m = \Xb_m
\hb_\gamma$. The $m$\th column of $\wt{\Gb}$ is then also denoted
by $\wt{\gb}_m = \Xb_m \mb_{H_\gamma}^{(r)}\in\RR^N$. Then,
$\Deltab_{ij}=\diag_M\cro{\wt{p}_{Q_{j}^m}^{(r-1)}(i)/v_{im}^{(r-1)}}$
and $\wt{\Hb}_{j}= \Espud{\wt{p}_{H_\gamma}^{(r)}}{\Gb\T
\Gammab_j^{(r-1)} \Gb}$ is an $M \times M$ matrix whose element
$(m,m')$ is given by:
\begin{align}
\Espud{\wt{p}_{H_\gamma}^{(r)}}{\gb_m\T \Gammab_j^{(r-1)} \gb_{m'}} &=
\Espud{\wt{p}_{H_\gamma}^{(r)}}{\gb_m}\T \Gammab_j^{(r-1)}
\Espud{\wt{p}_{H_\gamma}^{(r)}}{\gb_{m'}}+ \trace \bigpth{ \Gammab_j^{(r-1)} cov_{\wt{p}_{H_\gamma}^{(r)}}(\gb_m, \gb_{m'})} \nonumber\\
& = \wt{\gb}_m\T \Gammab_j^{(r-1)} \wt{\gb}_{m'} +
\trace\bigpth{\Gammab_j^{(r-1)} \Xb_m \Sigmab_{H_\gamma}^{(r)} \Xb\T_{m'}}.
\nonumber
\end{align}

\subsection{E-Q step:}\label{append:E-Q}
From $p(\Ab | \Qb)$ and $p(\Qb)$ in Section~\ref{JDEsec}, it follows that the $(\ab^m, \qb^m)$ couples correspond to
independent hidden Potts models with Gaussian class distributions.
It follows an approximation that factorizes over conditions:
$\wt{p}^{(r)}_Q(\Qb)\!=\!\!\prod\limits_{m=1}^M\wt{p}^{(r)}_{Q^m}(\qb^m)$
where $\wt{p}^{(r)}_{Q^m}(\qb^m) \!=\! f(\qb^m |
\ab^m\!=\!\mb_{A^m}^{(r)} ;
\mub_{m}^{(r-1)},\vb_{m}^{(r-1)},\beta_{m}^{(r-1)})$
   is the posterior of $\qb^m$ in a modified hidden Potts
model $f$, in which the observations $a_{j}^m$'s are replaced by
their mean values $\mb^{(r)}_{A_{j}^m}$ and an external field
 $\{\alphab_{j}^{m(r)}=v_{A_{j}^mA_{j}^m}^{(r)}
\cro{1/v^{(r-1)}_{{1m}},\ldots, 1/v^{(r-1)}_{{Im}}}\T, \quad  \linebreak j \in
\Pc_\gamma\} $ is added to the prior Potts model $p(\qb^m ;
\beta^{(r-1)}_m)$. It follows that the defined Potts reads:
%  so that it becomes:
\begin{align}
\label{eq:ising}
f(\qb^m ; \vb^{(r-1)}_m,\beta^{(r-1)}_m) \propto \exp \{ \sum_{j
\in \Pc_\gamma} \Big( \alphab_{j}^{m(r)}(q_{j}^m)
+\frac{1}{2}\beta^{(r-1)}_m \sum_{k \sim j} \scalarprod{q_{j}^m}{q_{k}^m}
\Big) \}.
\end{align}
Since the expression in Eq.~\eqref{eq:ising} is intractable, and
using the mean-field approximation~\cite{Celeux03},
$\wt{p}_{Q^m}^{(r)}(\qb^m)$ is approximated by a factorized
density $\wt{p}_{Q^m}^{(r)}(\qb^m) = \prod\limits_{j \in
\Pc_\gamma} \wt{p}_{Q_{j}^m}^{(r)}(q_{j}^m)$ such that if
$q_{j}^m=i$,
\begin{equation}
\wt{p}_{Q_{j}^m}^{(r)}(i)  \propto \Nc(\mb_{A_j^m}^{(r)} ;
\mu_{im}^{(r-1)} , v_{im}^{(r-1)})  f(q_{j}^m=i \I
\tilde{q}_{\sim j}^m; \beta_m^{(r-1)}, \vb_{m}^{(r-1)}),
\end{equation}
where $\tilde{\qb}^m$ is a particular configuration of $\qb^m$ updated
at each iteration according to a specific scheme and \linebreak 
$ f(q_{j}^m \I \tilde{q}_{\sim j}^m; \beta_m^{(r-1)}, \vb_{m}^{(r-1)}) \propto
\exp\{\alphab_{j}^{m(r)}(q_{j}^m)+\beta_m^{(r-1)} \sum\limits_{k
\sim j} \scalarprod{\tilde{q}_{k}^m}{q_{j}^m}\}$.

\subsection{M step:}\label{append:M}

% \begin{align}
%  \label{eq:M-4}
% \Thetab^{(r)}  &=
% \argmax_{\Thetab} \;  \Big[
% \Espud{\wt{p}_A^{(r)} \wt{p}_{H_\gamma}^{(r)}}{\log p(\Yb \I \Ab, \hb_\gamma, \Qb ; \Lb, \Gammab)} +
% \Espud{\wt{p}_A^{(r)} \wt{p}_Q^{(r)}}{\log p(\Ab\I \Qb ; \mub, \vb)}\nonumber \\
%  &+
% \Espud{\wt{p}_{H_\gamma}^{(r)}}{\log p(\hb_\gamma; \vb_h)} +
% \Espud{\wt{p}_Q^{(r)}}{\log p(\Qb ; \betab)}
% \Big].
% \end{align}

\subsubsection{M-$(\mub, \vb)$ step}\hfill \\
By maximizing with respect to $(\mub, \vb)$, Eq.~\eqref{eq:M-4} reads:
\begin{align}
%  \label{eq:M-4}
(\mub^{(r)}, \vb^{(r)})  &=
\argmax_{(\mub, \vb)} \;
\Espud{\wt{p}_A^{(r)} \wt{p}_Q^{(r)}}{\log p(\Ab\I \Qb ; \mub, \vb)}
\end{align}

\noindent By denoting $\bar{p}^{(r)}_{im}=\sum\limits_{j\in \Pc_\gamma}
\wt{p}^{(r)}_{Q_{j}^m}(i)$, and after deriving wrt $\mu_{im}$ and $v_{im}$ for every $i \in \{1\ldots I\}$ and
$m \in \{1\ldots M\}$, we get $\mu^{(r)}_{im} =\froc{\sum\limits_{j\in
\Pc_\gamma} p^{(r)}_{Q_{j}^m}(i)\; m^{(r)}_{A_{j}^m}}{\bar{p}^{(r)}_{im}}$ and
 $v^{(r)}_{im} = \froc{\sum\limits_{j\in
\Pc_\gamma} \wt{p}^{(r)}_{Q_{j}^m}(i) \bigpth{(m^{(r)}_{A_{j}^m }
- \mu^{(r)}_{im})^2 +
v^{(r)}_{A_{m}^jA_{m}^j}}}{\bar{p}^{(r)}_{im}}$.

% \begin{align}
%  \mu^{(r)}_{im} =\froc{\sum\limits_{j\in
% \Pc_\gamma} p^{(r)}_{Q_{j}^m}(i)\; m^{(r)}_{A_{j}^m
% }}{\bar{p}^{(r)}_{im}}&\;\rm{and}&
% v^{(r)}_{im} = \froc{\sum\limits_{j\in
% \Pc_\gamma} \wt{p}^{(r)}_{Q_{j}^m}(i) \bigpth{(m^{(r)}_{A_{j}^m }
% - \mu^{(r)}_{im})^2 +
% v^{(r)}_{A_{m}^jA_{m}^j}}}{\bar{p}^{(r)}_{im}}.
% \end{align}

% \begin{align}
%  \mu^{(r)}_{im} &=\froc{\sum\limits_{j\in
% \Pc_\gamma} p^{(r)}_{Q_{j}^m}(i)\; m^{(r)}_{A_{j}^m
% }}{\bar{p}^{(r)}_{im}} \nonumber \\
% v^{(r)}_{im} &= \froc{\sum\limits_{j\in
% \Pc_\gamma} \wt{p}^{(r)}_{Q_{j}^m}(i) \bigpth{(m^{(r)}_{A_{j}^m }
% - \mu^{(r)}_{im})^2 +
% v^{(r)}_{A_{m}^jA_{m}^j}}}{\bar{p}^{(r)}_{im}},
% \end{align}
% where $\bar{p}^{(r)}_{im}=\sum\limits_{j\in \Pc_\gamma}
% \wt{p}^{(r)}_{Q_{j}^m}(i)$.

%
% \begin{itemize}
%  \item $\mu^{(r)}_{im} =\froc{\sum\limits_{j\in
% \Pc_\gamma} p^{(r)}_{Q_{j}^m}(i)\; m^{(r)}_{A_{j}^m
% }}{\bar{p}^{(r)}_{im}}$,
% \item $v^{(r)}_{im} = \froc{\sum\limits_{j\in
% \Pc_\gamma} \wt{p}^{(r)}_{Q_{j}^m}(i) \bigpth{(m^{(r)}_{A_{j}^m }
% - \mu^{(r)}_{im})^2 +
% v^{(r)}_{A_{m}^jA_{m}^j}}}{\bar{p}^{(r)}_{im}}$.
% \end{itemize}
\subsubsection{M-$v_\hb$ step}\hfill \\
\label{par:v_hb}
By maximizing with respect to $v_\hb$, Eq.~\eqref{eq:M-4} reads:
\begin{align}
v_\hb^{(r)}  &=
\argmax_{v_\hb} \;
\Espud{\wt{p}_{H_\gamma}^{(r)}}{\log p(\hb_\gamma; \vb_h)}.
\end{align}

It follows the closed-form
expression given by
\begin{align}
v_\hb^{(r)} &= \frac{\Espud{\wt{p}_{H_\gamma}^{(r)}}{\hb_\gamma\T \Rb\M
\hb_{\gamma}}}{(D-1) } = \frac{(\mb_{H_\gamma}^{(r) t} \Rb\M
\mb_{H_\gamma}^{(r)}\! + \trace(\Sigmab_{H_\gamma}^{(r)}
\Rb\M))}{(D-1) }  = \frac{\trace((\Sigmab_{H_\gamma}^{(r)} + \mb_{H_\gamma}^{(r)}
\mb_{H_\gamma}^{(r) t}) \Rb^{-1})}{(D-1)}.
\end{align}

% \begin{itemize}
%  \item[] $v_\hb^{(r)} = {\displaystyle \frac{\Espud{\wt{p}_{H_\gamma}^{(r)}}{\hb_\gamma\T \Rb\M
% \hb_{\gamma}}}{(D-1) } \!= \!\frac{(\mb_{H_\gamma}^{(r) t} \Rb\M
% \mb_{H_\gamma}^{(r)}\! +\! \trace(\Sigmab_{H_\gamma}^{(r)}
% \Rb\M))}{(D-1) } = \frac{\trace((\Sigmab_{H_\gamma}^{(r)} +
% \mb_{H_\gamma}^{(r)} {\mb_{H_\gamma}}^{(r) t}) \Rb^{-1})}{(D-1)}}$.
% \end{itemize}

For a more accurate estimation of $v_\hb$, one may take advantage
of the flexibility of the VEM inference and inject some prior
knowledge about this parameter in the model. Since this parameter
is positive, a suitable prior can be an exponential distribution
with mean $\lambda_{v_\hb}^{-1}$:
\begin{equation}
\label{eq:priori_exp}
p(v_\hb;\lambda_{v_\hb}) = \lambda_{v_\hb} \exp\{-\lambda_{v_\hb}
v_\hb\}.
\end{equation}
Taking this prior into account, the new expression of the current
estimate $v_\hb^{(r)}$ is:
\begin{itemize}
 \item[] $v_\hb^{(r)} = \frac{(D-1) + \sqrt{8\lambda_{v_\hb}C + (D-1)^2}}{4\lambda_{v_\hb}}$
with $C = \trace((\Sigmab_{H_\gamma}^{(r)} + \mb_{H_\gamma}^{(r)}
{\mb^{(r)}_{H_\gamma}}^{t}) \Rb^{-1})$
\end{itemize}

\subsubsection{M-$\betab$ step}\hfill \\
By maximizing with respect to $\betab$, Eq.~\eqref{eq:M-4} reads:
\begin{align}
\betab^{(r)}  &=
\argmax_{\betab} \;
\Espud{\wt{p}_Q^{(r)}}{\log p(\Qb ; \betab)}.
\end{align}

Updating $\betab$ consists of making further use of a mean
field-like approximation~\cite{Celeux03}, which leads to a
function that can be optimized using  a gradient algorithm. To
avoid over-estimation of this key parameter for the spatial
regularisation, one can introduce for each $\beta_m$, some prior
knowledge $p(\beta_m;\lambda_{\beta_m})$ that penalises high
values. As in Eq.~\eqref{eq:priori_exp}, an exponential prior with
mean $\lambda_{\beta_m}^{-1}$ can be used. The expression to
optimize is then given by:
\begin{align}
\beta^{(r)}_m &= \arg\max\limits_{\beta_m}
E_{\wt{p}_{Q^{m}}^{(r)}} [\log
p(\qb^m ; \beta_m) p(\beta_m;\lambda_{\beta_m})]\nonumber \\
 &= \arg\max\limits_{\beta_m} \{-\log Z(\beta_m) +
\beta_m (\sum\limits_{j \thicksim k} \Espud{\wt{p}^{(r)}_{Q^m}}{
\scalarprod{q_{j}^m}{q_{k}^m}} - \lambda_{\beta_m}) \}.
\end{align}
After calculating the derivative wrt $\beta_m$, we retrieve the standard equation detailed in~\cite{Celeux03}
in which $\sum\limits_{j \thicksim k} \Espud{\wt{p}^{(r)}_{Q^m}}{
\scalarprod{q_{j}^m}{q_{k}^m}}$ is replaced by $\sum\limits_{j \thicksim k} \Espud{\wt{p}^{(r)}_{Q^m}}{
\scalarprod{q_{j}^m}{q_{k}^m}} - \lambda_{\beta_m}$. It can be easily seen that, as expected,
subtracting the constant $\lambda_{\beta_m}$ helps penalizing high $\beta_m$ values.

\subsubsection{M-$(\Lb, \Gammab)$ step}\hfill \\
This maximization problem factorizes over voxels so that for each
$j \in \Pc_\gamma$, we need to compute:
\begin{align}
(\ellb^{(r)}_j, \Gammab^{(r)}_j) &= \argmax\limits_{(\ellb_j,
\Gammab_j)} \;  \Espud{\wt{p}_{H_\gamma}^{(r)} \wt{p}_{A_j}^{(r)}}{\log
p(\yb_j \I \ab_j, \hb_\gamma; \ellb_j, \Gammab_j)}\;,
\end{align}
where $\ab_j = \{a_j^m, m=1 \ldots M\}$. Finding the maximizer wrt
$\ellb_j$ leads  to ($\wt{\Gb}$ is defined in the E-A step):
\begin{align}
\ellb^{(r)}_j &= \argmax\limits_{\ellb_j} \{2 (\wt{\Gb}
\mb_{A_j}^{(r)}
 - \yb_j)\T \Gammab^{(r)}_j \Pb \ellb_j  + \ellb_j\T \Pb\T\Gammab^{(r)}_j \Pb
\ellb_j \}.
\end{align}
\noindent After calculating the derivative wrt $\ellb_j$, we get
$\ellb^{(r)}_j= (\Pb\T \Gammab^{(r)}_j \Pb)\M\Pb\T \Gammab^{(r)}_j
(\yb_j -\wt{\Gb} \mb_{A_j}^{(r)})$.
In the AR(1) case, with $\Gammab_j =\sigma_j^{-2}\Lambdab_j$, we
can then derive the following relationship:
\begin{align}\label{eq:MaxLi}
\ellb^{(r)}_j &=(\Pb\T \Lambdab^{(r)}_j  \Pb)\M \Pb\T
\Lambdab^{(r)}_j \biggpth{y_j -
\wt{\Sb}_j m_{H_\gamma}^{(r)}}F_1(\rho^{(r)}_j),% \nonumber\\
% &=F_1(\rho^{(r)}_j),
\end{align}
where $F_1$ is a function linking the estimates $\ellb^{(r)}_j$
and $\rho^{(r)}_j$. The above formula is similar to that in
\cite[p.965, B.2]{Makni08}, when replacing $\hb_\gamma$ by
$m_{H_\gamma}^{(r)}$ and $\ab$ by $m_A^{(r)}$.

Denoting $\yb^{(r)}_j= \yb_j - \Pb \ellb^{(r)}_j$ and considering
the maximization wrt $\sigma^{2}_{j}$, similar calculations lead
to:
\begin{equation}\label{eq:MaxSigmaN}
\sigma^{2(r)}_{j} =\frac{1}{N} \left(
\Espud{\wt{p}_{A_j}^{(r)}}{\ab_j^{t} \wt{\Lambdab}_j^{(r)} \ab_j}
- 2 \mb_{A_j}^{(r)t} \wt{\Gb} \Lambdab^{(r)}_j \yb^{(r)}_j +
\yb^{(r)t}_j \Lambdab^{(r)}_j \yb^{(r)}_j\right) =
F_2(\rho^{(r)}_j, \ellb^{(r)}_j),
\end{equation}
where $F_2$ is a function linking the estimates
$\sigma^{2(r)}_{j}$ with $\ellb^{(r)}_j$ and $\rho^{(r)}_j$.
Matrix $\wt{\Lambdab}^{(r)}_j=
\Espud{\wt{p}_{H_\gamma}^{(r)}}{\Gb\T \Lambdab^{(r)}_j \Gb}$ is a
$M \times M$ matrix similar to the matrix $\wt{\Hb}_j$ introduced
in the E-A step. Its $(m,m')$ entry is given by
$\wt{\gb}_m\T \Lambdab_j^{(r)} \wt{\gb}_{m'} +
\trace\bigpth{\Lambdab_j^{(r)} \Xb_m \Sigmab_{H_\gamma}^{(r)}
\Xb\T_{m'}}.$\\
Eventually, the maximization wrt $\rho_j$ leads to
$\rho^{(r)}_j = \argmax\limits_{\rho_j} \{
\bigpth{\trace(\Ub_1\wt{\Lambda_j}) + \trace(\Ub_2
\Lambda_j)}/\sigma_j^{2(r)} +\log |\Lambda_j|\},$ with $|\Lambda_j|=
1-\rho_j^2$ and where $\wt{\Lambda_j}$ has the same expression as
$\wt{\Lambdab}^{(r)}_j$ without the $(r)$ superscript. Matrices
$\Ub_1$ and $\Ub_2$ are respectively $M \times M$ and $N \times N$
matrices defined as $\Ub_1=\Sigmab^{(r)}_{A_j} + \mb_{A_j}^{(r)}
{\mb_{A_j}^{(r)}}\T$ and $\Ub_2= \yb_j^{(r)}(\yb_j^{(r)} - 2
\wt{\Gb} \mb_{A_j}^{(r)})\T$. The derivative, denoted by $\Lambda_j$ of $\Lambda_j$  wrt
$\rho_j$ writes $\Lambda_j' = 2\rho_j \Bb+ \Cb$,  where
the entries of $\Bb$ and $\Cb$ are zero except $(\Bb)_{n,n}$ which
is 1 for $n=2:(N-1)$ and for $(\Cb)_{n,n+1}$ and $(\Cb)_{n+1,n}$
which are -1 for $n=1:(N-1)$. The derivative, denoted by $\wt{\Lambda}_j'$, of $\wt{\Lambda}_j$
wrt $\rho_j$ can be written as: $\wt{\Lambda}_j' = 2\rho_j
\wt{\Bb}+ \wt{\Cb}$ where $\wt{\Bb}$ and $\wt{\Cb}$ are $M \times
M$ matrices whose entries $(m,m')$ are respectively
 $(\wt{\Bb})_{m,m'} = \trace\bigpth{(\Xb_m \Sigmab_{\Hb_\gamma}^{(r)}
 \Xb_{m'}\T + \wt{\gb}_{m'}\wt{\gb}_m\T)\Bb}$
 and $(\wt{\Cb})_{m,m'} = \trace\bigpth{(\Xb_m \Sigmab_{\Hb_\gamma}^{(r)}
 \Xb_{m'}\T + \wt{\gb}_{m'}\wt{\gb}_m\T)\Cb}$.
Eventually, the derivative wrt $\rho_j$ leads to:
$$\rho_j^{(r)} = \frac{1-\rho_j^{(r)}}{\sigma_j^{2(r)}} \{2
\rho_j^{(r)} \bigpth{\trace(\Ub_1 \wt{\Bb}) + \trace(\Ub_2 \Bb)}+
\trace(\Ub_1 \wt{\Cb}) + \trace(\Ub_2 \Cb)\}= F_3(\rho_j^{(r)},
\sigma_j^{2(r)}).$$ Then $\rho_j^{(r)}$ can be
estimated as a solution of the fixed point equation
$\rho_j^{(r)}=F_3(\rho_j^{(r)}, F_2(\rho_j^{(r)},
F_1(\rho_j^{(r)}))).$
\vspace{.3cm} Note that
 in
the Gaussian noise case, the updating of the noise parameters
reduces to the estimation of $\sigma_{j}^{2(r)}$ which simplifies
into $\sigma_{j}^{2(r)} =\frac{1}{N} \left(
\Espud{\wt{p}_{A_j}^{(r)}}{\ab_j^{t}
\Espud{\wt{p}_{H_\gamma}^{(r)}}{\Gb\T \Gb} \ab_j} - 2
\mb_{A_j}^{(r)t} \wt{\Gb} \yb^{(r)}_j + \yb^{(r)t}_j
\yb^{(r)}_j\right)$.
% \begin{align}\label{eq:MaxSigmaNwhite}
% \sigma_{j}^{2(r)} &=\frac{1}{N} \left(
% \Espud{\wt{p}_{A_j}^{(r)}}{\ab_j^{t}
% \Espud{\wt{p}_{H_\gamma}^{(r)}}{\Gb\T \Gb} \ab_j} - 2
% \mb_{A_j}^{(r)t} \wt{\Gb} \yb^{(r)}_j + \yb^{(r)t}_j
% \yb^{(r)}_j\right) .
% \end{align}
% In the AR(1) case, an additional relationship can be found between
% $\rho^{(r)}_j$ and $\sigma^{2(r)}_j$ and this M-step can be
% performed by solving a fixed point equation.
%   {
% \bibliographystyle{plainnat}
{\footnotesize

\bibliographystyle{IEEEbib}
\bibliography{\BIBLIO}
% \bibliography{biblio}
 }

\end{document}